\documentstyle[12pt,aasms4]{article} 
\def\etal{{et~al.}\ }
\def\kpc{{\rm\,kpc}}

\def\kms{{\rm\,km/s}}

\def\msun{{\rm\,M_\odot}}

\def\vol#1  {{{#1}{\rm,}\ }}

\def\lya{{\rm Ly}\alpha}

\def\nhi{N_{HI}}

\def\etal{et al.\ }

\def\cf{{cf.}\ }

\def\clock{\count0=\time \divide\count0 by 60
     \count1=\count0 \multiply\count1 by -60 \advance\count1 by \time
     \number\count0:\ifnum\count1<10{0\number\count1}\else\number\count1\fi}

\begin{document}
\title{Sizes, Shapes, and Correlations of Lyman Alpha Clouds 
and Their Evolution in the CDM$+\Lambda$ Universe}
\author{Renyue Cen\altaffilmark{1,2} and Robert A. Simcoe\altaffilmark{1}}
\centerline{Email: (cen,rasimcoe)@astro.princeton.edu}

%\centerline{draft 4 of June 6, 1996}

\altaffiltext{1} {Princeton University Observatory, Princeton University, Princeton, NJ 08544}
\altaffiltext{2} {Department of Astronomy, University of Washington, Seattle, WA 98195}

\begin{abstract}
This study analyzes the 
sizes, shapes and correlations of $\lya$ clouds produced by a
hydrodynamic simulation
of a spatially flat 
CDM universe with a non-zero cosmological constant
($\Omega_0=0.4$, $\Lambda_0=0.6$, $\sigma_8=0.79$),
over the redshift range $2\le z \le 4$. 
The $\lya$ clouds range in size from
several kiloparsecs to about a hundred kiloparsecs in 
proper units, and they range in shape from 
roundish, high column density
 regions with
$\nhi\ge 10^{15}$~cm$^{-2}$
to low column density sheet-like structures 
with $\nhi \le 10^{13}$~cm$^{-2}$ at z=3.
The most common shape found in the simulation
resembles that of a flattened cigar.
The physical size of a typical cloud grows with time
roughly as $(1+z)^{-3/2}$ while its shape
hardly evolves (except for the most dense regions $\rho_{cut}>30$).
Our result indicates that any simple 
model with a population of 
spheres (or other shapes) of a 
uniform size is oversimplified;
if such a model agrees with
observational evidence, it is probably only by coincidence.
We also illustrate why the use of double quasar sightlines to 
set lower limits on cloud sizes is
useful only when
the perpendicular
sightline separation is small ($\Delta r \le 50h^{-1}$kpc).
Finally, we conjecture that high column density
$\lya$ clouds ($\nhi\ge 10^{15}$cm$^{-2}$)
may be the progenitors of the lower redshift faint blue galaxies. 
This seems plausible because their correlation length,
number density (extrapolated to lower redshift)
and their masses are in fair agreement with those observed.

\end{abstract}

\keywords{Cosmology: large-scale structure of Universe 
-- cosmology: theory
-- intergalactic medium 
-- quasars: absorption lines 
-- hydrodynamics}

\section{Introduction}

Acceptable theories of structure formation
are those whose parameters 
have been tuned 
to match the most up-to-date 
cosmological observations.  
At high redshift, the current leading constraint on models is the
data from COBE (\cite{s92}),
which fixes the amplitude
of the power spectrum on very large scales ($\sim 1000h^{-1}$Mpc)
to an accuracy of about 12\%.
At low (essentialy zero) redshift, we demand that models fit 
current observations of our local 
universe, primarily those concerning 
the distributions of galaxies in $(\vec x, \vec v)$ space.
These include the abundance of clusters of galaxies, 
which fixes the amplitude
of the power spectrum on scales of $\sim 8h^{-1}$Mpc
to about 10\% accuracy (\cite{bc92};
\cite{ob92};
\cite{bc93};
\cite{wef93};
\cite{vl95};
\cite{bm96};
\cite{ecf96};
Pen 1996),
the power spectrum of galaxies, which constrains the shape 
of the power spectrum on the intermediate-to-large scale of 
$\sim 10-100h^{-1}$Mpc
(Peacock \& Dodds 1994; Feldman, Kaiser \& Peacock 1994),
the ratio of gas to total matter in galaxy clusters, 
which determines $\Omega_b/\Omega_{tot}$ (\cite{wnef93}).
In addition,
the current measurements of the Hubble constant 
(Fukugita, Hogan, \& Peebles 1993;
Freedman \etal 1994;
\cite{rpk95};
Hamuy \etal 1995)
and the age constraint from the oldest globular clusters 
(Bolte \& Hogan 1995)
limit the range for $H_o$
and the combination of $\Omega_0$ and $\Lambda_0$.

This observational
suite has been examined by Ostriker \& Steinhardt (1995) 
to constrain flat cold dark matter models with a non-zero
cosmological constant
(\cite{p84};
\cite{ebw92};
\cite{bc92};
\cite{kgb93};
\cite{cgo93}).
The exercise is repeatable for other models (not necessarily
flat) including 
the tilted cold dark matter model 
(\cite{cgko92}; \cite{lls92}; \cite{lc92}; \cite{a93}; \cite{lmm93}),
the mixed dark matter model
(\cite{dss92}; \cite{tr92}; \cite{k93}; \cite{co94}; \cite{mb94}),
the open cold dark matter model (Gott 1982; Bucher, Goldhaber, \& Turok 1995)
and the primeval isocurvature baryon model (Peebles 1987a,b; 
Cen, Ostriker, \& Peebles 1993).
The allowed parameter space for 
the family of Gaussian cosmological models with the tunable parameters
($H_o$, $\Omega_0$, $\Lambda_0$, R$_{H/C}$, $n$, ISO, ADIA)
is thus quite limited.  In this context,
$H_o$, $\Omega_0$, and $\Lambda_0$ are the Hubble constant,
the density parameter, and the value of cosmological constant, all
in the present epoch.
R$_{H/C}$ is the mass ratio of hot to cold matter,
$n$ is the asymptotic power spectral index on large scales,
and ISO and ADIA denote isocurvature and adiabatic models.
Although it is possible to further tighten this parameter space
simply by making more accurate observations 
of the forementioned quantities, we should not overlook any
other {\em independent} tests which could help to
distinguish between contending models.

The $\lya$ forest lines observed in the 
spectra of high redshift quasars have two unique qualities
which distinguish them from other observationally accessible phenomena, 
at redshifts between the epoch observed by COBE
and our local universe.
First, each line of sight 
indiscriminately samples the
distribution of neutral hydrogen gas %in the intergalactic medium  
over a wide redshift range ($z\sim 0-5$)  
along random lines of sight (i.e., foreground objects are
unrelated to the background quasar).
Second, each individual spectrum contains 
a large amount of information about low absorption regions (e.g. voids,
fluctuating Gunn-Peterson absorption)
as well as information from high absorption ``cloud'' lines
(number densities of clouds at different redshifts, $b$-parameters
of the individual clouds, correlations of the clouds, 
relationship with other cosmic entities).
With a database of thousands of observed quasars, the total 
amount of information available is very large,
allowing for very detailed statistical studies
(e.g., Carswell \etal 1991; Rauch \etal 1993; Petitjean \etal 1993;
Schneider \etal 1993; Cristiani \etal 1995; Hu \etal 1995;
Tytler \etal 1995).  
These facts, taken together, suggest that the $\lya$
forests constitute perhaps the most rich and unbiased  
sample available for studying the universe at moderate redshift.

Recent cosmological hydrodynamic simulations 
by several independent
groups	have consistently shown that $\lya$ clouds 
are an integral part of the cosmic structure,
resulting naturally from the
gravitational growth and/or collapse of 
density fluctuations on
small-to-intermediate scales of $\sim 100$ kpc to a few Mpc
in comoving units (Cen \etal 1994, CMOR hereafter; 
Zhang \etal 1995; Hernquist \etal 1996).
Several simple population models
have been designed to examine the
(local) physical as well as (global) statistical properties
of $\lya$ clouds
(Sargent \etal 1980; Ostriker \& Ikeuchi 1983; Ikeuchi \& Ostriker
1986; Bahcall \& Spitzer 1969; Arons 1972; 
Rees 1986; Ikeuchi 1986;
Bond, Szalay, \& Silk 1988).
One of the essential simplifications in almost all these models
is to assume that individual $\lya$ clouds are spherical.
While the clouds produced in the new simulations may have some 
physical properties in common with these simple models,
a visual inspection of the new simulation 
results reveals that 
the $\lya$ absorbing structures resulting from
small-to-intermediate scale structure formation at high redshift
are {\it far from spherical}.
Furthermore, they seem to have a wide range of sizes. 
We presented these results first in CMOR,
and then in much more detail in Miralda-Escud\'e 
\etal (1996, MCOR hereafter).
Here, we present a quantitative study 
of the topological aspects of the $\lya$ clouds, 
complementing the topics covered in CMOR and MCOR.

In the redshift range from two to four,
$\lya$ clouds exhibit a rich 
spectrum of structure, ranging in shape from semi-spherical
to filamentary and even sheet-like. 
The spherical structures often
reside at density maxima (with $\nhi \ge 10^{15}$cm$^{-2}$)
located in the centers of extended
structures, while the filaments and sheets
tend to have a low $\nhi$ ($\le 10^{13}$cm$^{-2}$),
and form a web-like, interconnecting network covering 
large portion of the simulation box ($L=10h^{-1}$Mpc comoving).

We also show that it is not unusual for
pairs of quasar sightlines to contain absorption
features at close wavelengths. 
In particular,
we show that ``common'' absorption features 
(coincident lines) in sightline pairs
separated by $100h^{-1}$ proper kpc or more are probably 
due to absorption by {\it different clouds},
while on separations smaller than $40h^{-1}$kpc, double
sightlines are mostly likely to actually pierce a common
cloud.

Finally, we show that  $\lya$ clouds are spatially significantly
clustered, with a correlation length
of roughly $1-2h^{-1}$Mpc comoving (for the high column density
clouds of $\nhi\ge 10^{15}$cm$^{-2}$) at redshift $z\sim0.5-1.0$.
This suggests the possibility of an
intriguing connection to faint blue objects 
(\cite{k86}; \cite{t88}; \cite{clgm88}).

This paper is organized as follows.  Some brief descriptions of 
the simulations (for more details see MCOR)
and our cloud identification method are presented in \S 2.
Results and conclusions are given in \S 3 and \S 4.

\section{Simulations and Cloud Identification}

\subsection{Simulations}

We simulate the formation of $\lya$ clouds
in a spatially flat cold dark matter universe
with a cosmological constant ($\Lambda$CDM),
using the following cosmological parameters:
$H_o=65$km/s/Mpc, 
$\Omega_{0,CDM}=0.3645$, $\Lambda_0=0.6$,
$\Omega_{0,b}=0.0355$ (cf. Walker \etal 1991), 
$\sigma_8=0.79$
(the simulations we use in this paper
are the same as those used in CMOR and MCOR).
The primary motivation for choosing this model
is that it best fits the available observations which were
summarized in the introduction.
The simulation box size is $10h^{-1}$ comoving Mpc per side,
and contains $N=288^3$ cells and $144^3$ dark matter particles.
The cell size is $35~h^{-1}$ comoving kpc, corresponding
to a average baryonic cell mass of $6.3\times 10^5\msun$,
with the true spatial and mass resolutions being 
about 2 and 8 times worse than those values, respectively.
At $z=3$,
the Jeans length, $\lambda_J\equiv (\pi c_s^2/G\bar\rho_{tot})^{1/2}$
for $c_s=v_{rms}=10\kms$, 
is equal to $400 h^{-1} \kpc$ in comoving units, or 11 cells.
The power spectrum transfer function is computed
using the method described in Cen, Gnedin, \& Ostriker (1993).
We use a new shock-capturing 
Total Variation Diminishing (TVD) cosmological hydrodynamic code 
described in Ryu \etal (1993).

All the atomic processes for a plasma of (H, He) of primeval
composition (76\%,24\%) in mass are 
included, using the heating,
cooling, and ionization terms described in Cen (1992).
We calculate self-consistently
the average background photoionizing
radiation field as a function of frequency, assuming the radiation field
is spatially uniform (i.e., optically thin).
 The evolution
of the radiation field is calculated given the average attenuation in the
simulated box and the emission (both from the gas itself and from the
assumed sources of ionizing photons).
The time-dependent equations for the ionization structure of the gas
are solved by iteration using an implicit method, to avoid the
instabilities that arise in solving stiff equations. 
In general, the
abundances of different species are close to ionization equilibrium 
between recombination and photoionization
after most of the gas has been photoionized.

We model galaxy formation as in Cen \& Ostriker (1992, 1993a,b).
The material turning into collisionless particles as 
``galaxies" is assumed to emit ionizing radiation, with two types of spectra:
one characteristic of star formation regions and the other
characteristic of quasars, with efficiencies (i.e., the fraction of
rest-mass energy converted into radiation) of
$e_{UV,*}=5\times 10^{-6}$, and $e_{UV,Q}=6\times 10^{-6}$, respectively.
We adopt the emission spectrum of massive stars from Scalo (1986) 
and that of quasars from Edelson and Malkan (1986).
Details of how we identify galaxy formation and
follow the motions of formed galaxies
have been described in Cen \& Ostriker (1993a).
Note that in this simulation
supernova energy feedback into the intergalactic medium 
from aging massive stars is not included.

\subsection{$\lya$ Cloud Identification}

Perhaps the most critical decision which must be made in a study such
as this is how to define a $\lya$ ``cloud''.  
This is not as simple as
it may seem, because the smooth transition in density 
between the global intergalactic medium and the
local structures within it can blur
the distinction between actual clouds and the intercloud medium.  
However, in order to examine the structure of  $\lya$ clouds
in any quantitative way, it is absolutely necessary to adopt some sort
of definition of the boundary of a cloud.  
The simplest approach which we believe to be meaningful is to 
identify clouds as regions with densities above a
chosen threshold. 

There are two major motivations
for defining our clouds in this manner. 
First of all, this method allows us to associate the clouds
found at each density cut with a particular column density,
once we obtain information about the 
characteristic sizes of the clouds.    
This is a desirable feature
because we can only observe $\lya$ forest
clouds through absorption lines in QSO spectra. 
Second, density perturbations with
amplitudes larger than a certain threshold become
self-gravitating, bound clouds.  
For a spherical cloud,
the average overdensity at which a cloud breaks
away from the general Hubble expansion and 
becomes self-gravitating is $\sim5.55$.
Thus, our cloud defintion also has physical motivation in
that structures above some high density threshold 
are gravitationally bound, distinct systems. 
Having chosen a general strategy for defining our clouds, we now
describe the details of the grouping procedure.

Cells with a baryonic density below a chosen value 
(3,10, and 30 will be used)
in units of the global mean of the baryonic density
are cut out of the original density array. 
Then, clouds are defined by grouping
the remaining cells using the
DENMAX scheme (\cite{bg91}) as follows.

First, two sets of baryonic densities are stored.
One is the original density output array 
from the simulation ($\rho_{org}$, $288^3$ elements), and 
the second is a smoothed version of $\rho_{org}$
with a Gaussian smoothing window of radius 1.5 cells
($\rho_{smooth}$, $288^3$ elements).
The smoothing is performed to eliminate any 
small, cell-to-cell density fluctuations, which could be physical
in the form of small discontinuities or oscillating sound waves, or
just numerical noises. 
Such small-scale fluctuations could result in
the identification of small unwanted clouds 
with sizes on the order of one cell.
At the same time, 
since the true resolution of the simulation is close to 2-3 cells,
a Gaussian smoothing window of radius 1.5 cells 
should preserve all information regarding ``real'' adjacent structures.
A larger smoothing window could
merge real and separate entities which are actually
well-resolved in the simulation.
We have made rather extensive experiments on the smoothing operation
and conclude that the adopted smoothing window size is appropriate,
as will be shown below.

After the smoothing operation is performed,
we cut out all elements of $\rho_{smooth}$ 
and $\rho_{org}$
whose $\rho_{org}$ values are below a chosen value, $\rho_{cut}$.
We then collect the uncut cells of $\rho_{org}$ 
(using the DENMAX scheme) by propagating each cell
along the gradient of $\rho_{smooth}$
until it reaches a local density maximum.
All the cells collected at a particular local maximum are
grouped into one ``cloud".

Once the cells have been grouped, we examine the shapes and sizes of 
the resultant ``clouds" by 
approximating them as {\it ellipsoids}.  
A symmetric $3\times3$ moment tensor is created for each cloud 
of the form $T_{11}=\Delta x^2$; $ T_{22}=\Delta y^2$; 
 $T_{12}=\Delta x \Delta y;$  and so on.  
This tensor is diagonalized,
and the square roots of the eigenvalues yield the lengths of the 
semi-major and semi-minor axes,
$a^\prime,b^\prime,c^\prime$, in descending order of length.
For cells where $\Delta x$, $\Delta y,$ 
or $\Delta z=0$, we account for the finite size of a single cell 
by finding ${\int x^2 dydx \over \int dydx} = \frac{1}{12}$
for a 1 cell square and adding that to 
the tensor component for a particular cloud.
This method underestimates the actual axial lengths of an ellipsoid,
because it only computes moments.
We attempt to correct this by scaling up the values of 
$a^\prime,b^\prime,$ and $c^\prime$
 (determined above)
by a factor of 
$\left( {V \over 4\pi a^\prime b^\prime c^\prime /3}\right) ^{1/3}$: 
\begin{equation}
(a,b,c) = (a^\prime,b^\prime,c^\prime) ({V \over 4\pi a^\prime b^\prime c^\prime /3})^{1/3}, 
\end{equation}
where $V$ is the actual volume of the cloud 
computed by summing over all the member cells of the cloud.

\section{Results}

\subsection{A Visual Inspection}

We first present a three dimensional visual description of the
structure of the $\lya$ clouds produced in the simulation.
Figures (1a,b,c) display the isodensity contour surfaces at $z=3$
for $\rho_{cut}=(3,10,30)$, respectively.
Here, $\rho_{cut}$ is the baryonic density in units of its global mean.
The box size is $10h^{-1}$ comoving Mpc.
The most striking visual feature is the 
network of sheets and filaments spanning the box.
In Figure 1a we see that
most of the structures are connected over
the entire box, and it is apparent that
most of the covering areas [i.e., cross section for quasar lines 
of sight at low column densities ($\le 10^{13}$~cm$^{-2}$; see below)]
are sheet-like objects.
Although this result is for the particular $\Lambda$CDM universe
simulated here, we expect this qualitative picture to be fairly generic
for any CDM-like model (HDM-like models should show
even more prominent sheets,
while models like PBI are likely to be less coherent).
At a higher density of $\rho_{cut}=10$ (Figure 1b) 
most structures become
filamentary and more isolated.
When the isodensity reaches  $\rho_{cut}=30$ (Figure 1c),
most of the filaments are replaced by relatively 
round systems, which are typically isolated with separations 
of about one comoving megaparsec.

The second noticeable visual feature 
is the existence of a large, low density 
region at the center of the box,
occupying more than half of the simulation volume.
Since the initial condition of the simulation 
is not constrained (i.e., it is only a random realization of the 
cosmological model in such a volume),
this void indicates that the simulation box may
still not be large enough
to contain a ``fair" volume for the structures under consideration.
It may be necessary
to use a larger simulation box,
perhaps $20-30h^{-1}$Mpc on a side, in order to 
properly sample the objects in question.
Different properties will be
affected by the simulation volume to varied degrees.
We expect that the most significant differences will
be in quantities like void sizes, or correlations.

Let us now examine the individual clouds  
to assess the general accuracy of our grouping scheme.
Figure 2a shows clouds at $\rho_{cut}=3$
for two randomly selected
slices of size $5\times 5h^{-2}$Mpc$^2$
with thickness of $175h^{-1}$kpc (all lengths are in comoving units)
at $z=3$.
The two left panels show the density contours of levels
$10^{(i-1)/2}~\rho_{cut}, i=1,2,3...$, and 
the two right panels show the identified $\lya$ clouds
in the same slices.
Two symbols are used to show two types of clouds:
the filled dots represent ``large" clouds with the 
dot size roughly
proportional to the actual size of the cloud, and
the open circles are for the ``small" clouds.
A cloud is called ``small" if its smallest axis 
does not exceed 3 cells in length, and ``large" if it does.
Figure 2b is similar to Figure 2a,
but with $\rho_{cut}=10$, and for two different slices.
We note that
``large" clouds are typically embedded in larger, 
extended density structures
while ``small" ones are more isolated.
A close, one-by-one examination of the density maxima in
the density contour plots indicates
that each identified cloud corresponds
to a well defined density maximum
(Some clouds, especially ``small" ones,
have no corresponding density maxima simply because
their densities fall below the contour levels when averaged over
the slice thickness to be displayed). 
It appears that our smoothing operation and
cloud identification scheme
indeed yield well defined, distinct clouds with one particularly
desirable feature:  
they are neither over-merged in the sense
that a single cloud contains multiple structures of comparable size
(over-smoothed), nor over-separated in the sense that artificial
clouds are created by small-scale small-amplitude density
fluctuations (undersmoothed).

When the clouds are classified into ``large" and ``small" groups,
we find that most of the clouds are ``small": 
the fraction of
``small" clouds at $\rho_{cut}=(3,10,30)$ at $z=3$ is
%($42\%, 74\%, 93\%$) 
($62\%, 92\%, 98\%$), 
containing
%($9\%, 38\%, 51\%$) 
($14\%, 49\%, 63\%$) 
of the total cloud mass.
The fact that a large fraction of the baryonic mass 
is in ``small" clouds indicates that a
{\it higher resolution simulation is perhaps needed}
before we can be absolutely sure that 
these clouds are resolved properly.
Inclusion of more initial small scale power, which
is limited by the Nyquist frequency of a simulation,
might further increase the fraction of ``small" clouds.

We note in passing that we also experimented with the
conventional friends-of-friends grouping algorithm, with
a linking length of one cell.
The resulting clouds were obviously not useful for our analyses.
For example, at $\rho_{cut}=3$, most of the 
volume, as well as mass, wound up in a single supercloud 
because the distinct clouds as shown in Figure 2 were linked together
by touching boundaries.
In other words, the friends-of-friends grouping scheme
picks out large, inter-connected networks (see Figure 1a)
but not the individual structures embedded within them.

\subsection{Quantitative Measures}

We have chosen to focus on three quantitative aspects of the
identified $\lya$ clouds:
their sizes, shapes, and correlations.
In addition, double quasar
sightline analysis is performed.

To facilitate quantitative discussions,
we relate the mass density of a cloud 
to its column density.
Assuming that the clouds are in photoionization equilibrium,
the column density of a cloud
can be related to the parallel size and density of the cloud 
(assuming that the density is uniform across the cloud along 
the line of sight) as
\begin{equation}
N_{H} = 8.18\times 10^{11} ({\Omega_b\over 0.0125h^{-2}})^2({j_{HI}\over 10^{-12}})^{-1}({T\over 10^4})^{-0.7}{L\over 100 kpc}({\rho_{cut}\over \langle\rho_{b}\rangle})^2 ({1+z\over 4})^6~\hbox{cm}^{-2} ,
\end{equation}
where $\Omega_b$ is the baryonic density parameter,
$T$ the temperature in Kelvin,
$j_{HI}$ the hydrogen photoionization rate,
$\rho_b$ and $\langle\rho_b\rangle$ the baryonic 
density of the cloud and 
the mean baryonic density at the redshift in question,
$L$ the proper size of the cloud along the line of sight,
and $z$ the redshift.
For a photoionization radiation field 
with power-law form $\nu^{-1}$,
$j_{HI}=4.34\times 10^{-12}J_{LL}$~sec$^{-1}$,
where $J_{LL}$ is the  intensity of the 
photoionization field at the Lyman limit
in the usual units ($10^{-21}$erg/cm$^2$/hz/sec/sr).
Taking
$\Omega_b=0.0125h^{-2}$,
$j_{HI}=7.0\times 10^{-13}$sec$^{-1}$,
$T=2\times 10^4$~Kelvin, 
$L=40$~kpc as typical values at $z=3$,
one finds that 
$\nhi=(2.5\times 10^{12}, 2.7\times 10^{13}, 2.5\times 10^{14})~\hbox{cm}^{-2}$
for $\rho_{cut}=(3,10,30)$.
Of course, exact column densities would depend on exact values
of the assumed quantities as well as the actual density distribution.
We expect that the actual density distribution, which is non-uniform,
will increase the column densities. 
For example,
redistribution of the same mass within $L$ into a
(one dimensional) coreless, non-uniform one would result
in a column density 
$(1+\alpha)^2/(1+2\alpha)$
times that of the uniform density distribution,
where $\alpha$ is the slope of the density profile.
We simply increase the above estimates by a factor of four to
account for the gradient in the density distribution of a cloud
[note that this factor of four is fairly plausible for 
a reasonable $\alpha$ ($\sim -1/2$) in the one dimensional
singular case or a relatively lower $\alpha$ value for a 
non-singular profile.
We also note 
that this factor of four can also be achieved if $\Omega_b$ is
twice the value adopted here, in the light of recent new measurements
on the deuterium abundance (\cite{tfb96})],
thus obtaining
\begin{equation}
\nhi=(1.0\times 10^{13}, 1.1\times 10^{14}, 1.0\times 10^{15})~\hbox{cm}^{-2}
~~~\hbox{for}~~~\rho_{cut}=(3,10,30)~~~\hbox{at}~~z=3.
\end{equation}
\noindent We will use the above relation for quantitative discussions.

\subsubsection{A Few Simple Global Quantities}

Before engaging in a more detailed quantitative discussion
of cloud
properties, we present a few simple, global quantities.
Listed in Table 1 are the numbers of clouds at each  
redshift and density threshold,
the baryonic mass fraction of such clouds,
the mean inter-cloud separation in proper units,
and the mean cloud mass.
Two points are interesting.
First, while the average cloud increases its mass by a factor 
of about $3-4$ from $z=4$ to $z=2$,
the number of clouds decreases by a factor two.
This indicates that the merging of old clouds and 
creation of new clouds
are of comparabe importance.
Second, the fraction of baryonic mass contained in
$\lya$ clouds is large.
For example, at redshifts (2,3,4), (31\%, 29\%, 26\%) of baryons
are found in regions with densities between
 $\rho_{cut}=3$ and
 $\rho_{cut}=30$ (i.e., $\nhi\sim 10^{13} - 10^{15}$cm$^{-2}$).

\subsubsection{Sizes of $\lya$ Clouds}

We define the characteristic ``size" of a  cloud as

\begin{equation}
S\equiv (abc)^{1/3}, 
\end{equation}

\noindent where $a$, $b$, and $c$ are the semi-axes
of the ellipsoid approximating the cloud, measured
in proper physical units (see \S 2.2 for definition of $a,b,c$).
For a true ellipsoid cloud, the volume is $V={4\pi\over 3} S^3$.

Figure 3 shows the cumulative distribution of
cloud sizes at $z=3$ for $\rho_{cut}=(3,10,30)$, respectively. 
The thin curves are weighted by $A$, where 
\begin{equation}
A\equiv ab + ac + bc ,
\end{equation}
\noindent and the thick curves are weighted by mass. 
 	 $A$ is an approximation of the average covering area of 
a cloud on the sky,
which is proportional to the probability of the cloud 
intersecting a line of sight.
The $A$-weighted and mass-weighted distributions are very
similar, implying that $A$ and mass are correlated.
Clouds are larger at lower $\rho_{cut}$, as expected,
and they exhibit a wide range in size,
from a few to about a hundred proper kiloparsecs.
The median size varies from $15h^{-1}$~kpc at $\rho_{cut}= 30$
to $35h^{-1}$~kpc
for $\rho_{cut}= 3$.
Note that $S$ is only the cube root of the effective
volume of a cloud, so the distribution of cloud sizes
would be more varied,
if we used $a$ 
(semi-major axis of an ellipsoid) instead of $S$
as a measure of size.

Figure 4 shows the redshift evolution of the median
size of the $\lya$ clouds. 
Note that here the size (vertical axis) is 
in comoving length units, which might be easier to interpret.
%In a system of isolated, non-interacting clouds, one would expect
%the comoving sizes of high density clouds to shrink faster than those
%of low density clouds, because regions with more gravitating matter
%should collapse faster.  
Our data indicates that $\lya$ clouds grow with time
in comoving length units at a moderate rate, and that
this rate depends {\it weakly} on the density (i.e., three curves
for the three density cases are almost parallel to each other).
The fact that denser regions are not shrinking, but rather
expanding at a rate similar to that of less dense regions implies
that the merger/accreation rate 
is {\it higher} in the dense regions than in less dense regions.
Without merger/accretion effects, the high density
regions would appear to shrink due to gravitational collapse. 

To summarize, we find that $\lya$ clouds exhibit a wide range 
of sizes from a few proper kiloparsecs to about
a hundred proper kiloparsecs at z=3.
The comoving size of these clouds tends to increase 
slowly with time, implying
that the proper physical size increases with time more rapidly
than $(1+z)^{-1}$.

\subsubsection{Shapes of $\lya$ Clouds}

Figures (5a,b,c) show the 
distributions of cloud shapes
in the $c/b$~-~$b/a$ plane for clouds at $\rho_{cut}=(3,10,30)$
at $z=3$. 
Each dot represents one cloud and the contours 
indicate the density of clouds (weighted by $A$) 
in the plane.
The contour levels are incremented up linearly from outside to inside.
We see that the most common cloud at $\rho_{cut}=3$ and $10$
is an ellipsoid with axial ratios of $\sim 1:2:4$.
However, at $\rho=30$ the situation is interestingly different.
Two major concentrations of clouds 
are seen at $(b/a,c/b)=(0.35,0.85)$ and $(0.75,0.85)$,
indicating the existence of two distinct 
populations: filaments and near-spherical ellipsoids.

In order to conveniently show the evolution of shapes of the clouds 
and to relate shapes to other quantities,
we define a simple ``shape" parameter of a cloud as

\begin{equation}
\eta \equiv {b^2 + c^2\over a^2}. 
\end{equation}

\noindent
For an ideal (spherical, filamentary, disk) cloud,
$\eta=(2,0,1)$.  
Some degeneracy exists for any value of $\eta$ with such a simple
definition.
For example, $(a,b,c)=(1,1,0)$ and 
$(a,b,c)=(1,1/\sqrt{2},1/\sqrt{2})$ both give $\eta=1$.

Figure 6 shows $\eta$ as a function of size $S$ at $z=3$
for $\rho_{cut}=(3,10,30)$.
There is a trend, albeit with a large scatter, that 
larger clouds tend to be less spherical.

Figure 7 represents the $A$-weighted
cumulative distribution of $\eta$ for three cases at $z=3$.
We note two points.
First, we see that there is 
negligibly small fraction of clouds with the shape parameter $\eta$ 
greater 1.5, indicating that spherical 
clouds ($\eta=2$) contribute only a tiny percentage to the total
covering area of the clouds at all three density cuts. 
Second, we find a larger fraction of round clouds 
(large $\eta$) {\it and} filamentary clouds ($\eta$ close to zero)
at high densities than at a lower densities.
This is due to the fact that the large sheets 
and filaments present at lower
densities are broken up into both filaments and 
turn into roundish regions at higher densities, respectively.
This quantitative result is consistent with the visual impression
that structures go from sheets to filaments and 
spheres as the density goes from low to high (Figure 1).
The fact that there are more filaments than spheres at all densities
is probably a consequence of nonlinear evolution coupled with
hydrodynamic (pressure) effects.
The situation for dark halos might be different,
since dark matter
distributions tend to be more unstable in lower dimensional structures.

Finally, we show in Figure 8
the redshift evolution of the median shape paramter $\eta_{med}$.
It indicates that at $\rho_{cut}=3$
the median shape of the clouds do not change with time,
while at $\rho_{cut}=30$ 
the clouds progress toward the shape of spheres 
(larger $\eta$)
in time with 
a dramatic upturn from $z=3$ to $z=2$,
possibly related to the collapse/merging
of gas along the longest axis.
The situation at the intermediate $\rho_{cut}=10$
is between the above two cases.

In summary, we have shown that $\lya$ clouds display
a variety of shapes ranging from semi-spherical clouds
to filaments and sheets, with the most common shape 
being a ``flattened cigar" with an axial ratio of $\sim 1:2:4$
at column density of $10^{13}-10^{14}$ cm$^{-2}$,
and being either a thin cigar with an axial ratios of
$(1:3:3.4)$ or near sphere with an axial ratios of $(1:1.3:1.5)$
at column density of $>10^{15}$cm$^{-2}$.
Larger clouds tend to be less spherical.
The shapes of clouds with densities a few times the mean density
of the universe evolve very weakly with time,
whereas the clouds at higher densities ($\rho_{cut}\ge 30$)
grow more spherical with time.

\subsubsection{Double Quasar Line of Sight Analysis}

None of our analyses so far have been 
directly comparable with observables.
Here we attempt to project our theoretical 
models onto the observational plane.
CMOR and MCOR have examined many of the 
properties of these clouds in detail,
and found that the simulation results match observations fairly well.
Here, we focus on the observations of double quasar sightlines 
(\cite{sssfcwwm92};
\cite{d94};
\cite{d95};
\cite{srsrwk95};
\cite{fdcb96};
\cite{by96}).
Let us first describe our procedure for modeling these
observations.

1. Once three dimensional clouds in the simulation box
 are identified (see \S 2.2),
each cell is labelled with an integer $n$,
meaning that it belongs to cloud $n$ (if $n$ is zero, it means
that the cell does not belong to any cloud, i.e., its density 
is below $\rho_{cut}$).

2. A random pair of sightlines separated by $\Delta r$ 
is selected. 
Since the simulation box has no preferred orientation, 
we simply choose a direction perpendicular 
to one of three faces of the simulation box as
the direction for the double sightlines.

3. Along each of the two sightlines we identify cells
whose labels $n$ (see Step 1) are non-zero, and then separate regions
of different $n$ into separate clouds, 
which we call Clouds-Along-Line-Of-Sight
(CALOS's). Note that each CALOS inherits its 3-d parent cloud
label $n$. 

4. We define the location of each CALOS in $z$-space by finding the 
$\rho_{gas}^2$ weighted center of the CALOS cells located
along the sightline.  

5. For each CALOS along sightline \#1,
if a CALOS within a velocity space separation $D$
along the sightline \#2 is found, the following
operation is performed:
an integer counter $N_{co, common}$ is 
incremented if the two CALOS's share the same $n$, i.e.,
they belong to the same 3-d cloud.
Otherwise an integer counter $N_{co,clustering}$ is 
incremented by $1$.
If both a common and a clustering line are found, we consider
it to be a common pairing.
The above exercise is repeated for
each CALOS along sightline \#2.
We note that the bulk peculiar velocity of each CALOS is ignored
in this calculation, i.e., 
the line-of-sight velocity difference between
a pair of coincident lines simply reflects the 
line-of-sight real space distance difference.
This should be a good approximation, because the peculiar velocity
gradient is typically smaller than the Hubble constant 
on the scale $D$ (see below) in which we are interested.

We perform this exercise on 10,000 
randomly selected pairs of sightlines
for each value of $\rho_{cut}=(3,10,30)$ at 
six different perpendicular separations, 
$\Delta r=(1,3,6,12,24,48)$~cells.
Figure 9 shows the fraction of lines which are 
coincident at $z=3$, $N_{co,tot}/N_{tot}$,
as a function of $\Delta r$,
where $N_{co,tot} = N_{co,common} + N_{co, clustering}$
and $N_{tot}$ is total number of lines.
Six curves are shown, corresponding to three values of 
$\rho_{cut}=(3,10,30)$, each at two values of $D=(50,150)$~km/s.
We see that 
{\it the line coincident rate decreases with increasing
column density}.
At a perpendicular proper separation of $100h^{-1}$kpc and
$\rho_{cut}=3$ 
(corresponding to clouds with 
$\nhi=10^{13}$cm$^{-2}$) 
about $1/2$ to $3/4$ of the sightlines should have
coincident absorption features for $D=50$km/s to $150$km/s.
The line coincident rates drop to $(27\%,38\%)$ and $(11\%,17\%)$
for $\rho_{cut}=10$ and $\rho_{cut}=30$
(corresponding to clouds with 
$\nhi=10^{14}$cm$^{-2}$ and $\nhi=10^{15}$cm$^{-2}$), 
respectively, 
at $D=(50,150)$km/s.

Figure 10 shows the redshift evolution 
of $N_{co,tot}/N_{tot}$ at $\Delta r=100h^{-1}$~kpc in proper units,
for four cases with $(\nhi,D)$ of
$(3\times 10^{13},150)$,
$(3\times 10^{13},50)$,
$(3\times 10^{14},150)$,
$(3\times 10^{14},50)$.
The indicated neutral hydrogen column densities
in Figure 9 are approximately computed as follows.
Combining Equation (2) for $N_{HI}$ at $z=3$
(with the adopted fiducial values of various quantities)
with the result that the physical sizes of clouds
go roughly as $(1+z)^{-3/2}$ (Figure 4) gives
$\nhi(\rho_{cut},z)=1.0\times 10^{13}({\rho_b\over 3.0})^2 (1+z)^{9/2}$cm$^{-2}$.
We see that, {\it at a fixed column density}, 
the rate of coincident lines (at $\Delta r=100$~km/s)
{\it increases with redshift at a moderate pace},
and higher column density clouds and/or smaller 
$D$ yield weaker evolution,
assuming that $j_{HI}$ is {\it constant} 
over the redshift range in question.
The assumption of constant $j_{HI}$ is merely for the 
convenience of illustration. In fact, the self-consistently
produced photoionization field during the simulation 
is constant from $z=4$ to $z=2$ to within a factor of two.
In other words, the results would not have
significantly differed if the actual photoionization field 
had been used.

Given the three dimensional nature of the simulations,
we can distinguish between two kinds of coincident lines:
one of which ($N_{co, common}$)
occurs when a pair shares the same 3-d parent cloud,
the other of which ($N_{co, clustering}$)
occurs when the lines intersect two separate 3-d clouds.
The former set of coincident lines is due to the extended size
of a single cloud, while the latter set of coincident 
lines is due to
the clustering of several clouds.
Figure 11 shows $N_{co,common}/N_{co,tot}$ at $z=3$.
Six curves are shown 
for $\rho_{cut}=(3,10,30)$ with $D=50$~km/s and $150$~km/s.
We see that at $\Delta r\approx 30-60h^{-1}$kpc about half
of the coincident line pairs share the same clouds and
the other half pierce separate clouds.
At $\Delta r=100h^{-1}$kpc only 10-20\%
of the total coincident line pairs share the same clouds,
i.e., clustering of separate clouds dominates the coincident 
events at this and larger separations.

The observational signature of 
clustering dominated coincident lines 
is that the difference in line of sight velocity  
between the two absorption lines
should be a weak function of the  
separation perpendicular to the line of sight.
Figure 12 shows 
rms velocity difference of coincident line pairs along the
line of sight, $R_{||,rms}$, as a function of the
perpendicular separation, $\Delta r$,
 for $D=50$~km/s and 150km/s at $\rho_{cut}=(3,10,30)$. 
The two long-dashed horizontal lines show the rms difference
for the two $D$ values, if clouds are small (compared 
to perpendicular separation) and randomly distributed (i.e., no clustering).
We see that $R_{||,rms}$ depends rather weakly on 
$\Delta r$ when $\Delta r \ge 200h^{-1}$kpc, 
and it approaches the asymptotic limit of a random 
distribution of clouds at $\Delta r\approx 400-500h^{-1}$kpc.
We predict that for quasar double sightlines at large separations
($\Delta r>500h^{-1}$kpc), the rms velocity
difference between absorption features common to both sightlines,
$R_{||,rms}$, should depend on the
range being searched in $z$-space, $D$, as
\begin{equation}
R_{||,rms} = 0.58 D .
\end{equation}
 
Figures (13a,b) show the redshift evolution of $N_{co, common}/N_{co,tot}$
for $\nhi=3\times 10^{13}$cm$^{-2}$
and $\nhi=3\times 10^{14}$cm$^{-2}$, respectively.
Four cases are shown in each figure
with two separations, $\Delta r=50h^{-1}$kpc and 
$\Delta r=100h^{-1}$kpc both in proper units, and
with $D=50$km/s and $150$km/s.
We see that the fraction of coincident lines 
with a separation of $50-100h^{-1}$ proper kpc 
due to common clouds
{\it decreases with increasing redshift
at fixed column density}, 
presumably because 
clouds are smaller and more crowded at high redshift
than at low redshift.
We remind the reader that the total line coincident
rate {\it increases with redshift} (see Figure 10),
indicates that a higher cloud number density (crowding of clouds)
at high redshift dominates the line coincident events.
This is consistent with results shown in Figure 11.

The combined results of our quasar
double sightline analysis suggest a few
basic trends which 
should be kept in mind when analyzing QSO spectra.
The most important of these is that line coincident events can be
caused by two different phenomena.
At small perpendicular separations ($\Delta r < 40h^{-1}$kpc proper)
a pair of 
coincident lines is most likely to pierce a common cloud,
as is usually assumed when analyzing such observations to infer
the actual cloud size.
However, at larger $\Delta r$, a pair of 
coincident lines is more likely to penetrate two different clouds
which are spatially clustered.
At $\Delta r=100h^{-1}$ kpc, only 10-20\%
of the total coincident line pairs share the same clouds.
And at very large separations of
$\Delta r>500h^{-1}$kpc,
the coincident line events can be explained as the 
random intersection of two unrelated clouds,
whose rms difference $R_{||,rms}$ should relate to the velocity interval
$D$ as indicated in Equation (7).

\subsubsection{Correlations of $\lya$ Clouds}

It is tempting to make a connection between $\lya$ clouds
and dwarf galaxies and/or moderate redshift faint blue galaxies.
We approach this problem by examining clustering properties of
the $\lya$ clouds.

A few cautionary words about the limitations of the simulation 
are in order here.
The simulation box size ($L=10h^{-1}$Mpc comoving)
places severe limits on our ability to study the clustering properties
of clouds on large scales with a high degree of accuracy, because it
sets an upper limit on the scale of the input power spectrum.
Waves longer than the simulation box size
would have made a considerable contribution
to clustering on larger scales, since the density fluctuations on scales
comparable to the simulation box length have started to
approach nonlinearity
even at redshift $z=2$ 
($\sigma\sim 0.4$ at $z=2$ on the box scale).
The situation becomes even more serious at lower redshift.

Nevertheless, we may gain some insight into the distribution of mass
on smaller scales, and say somthing useful about the more local
clustering properties of clouds. 
The three dimensional
correlation of the $\lya$ clouds at $z=3$ is shown
in Figure 14 for $\rho_{cut}=(10,30)$.
Higher column density clouds seem to be
more strongly clustered than lower column density ones.
Since most of the faint blue galaxies are 
thought to be in the redshift $z\sim 0.5-1.0$, we
show in Figure 15 the evolution of 
correlation length $r_o$ (solid curve).
This is defined as the length where the correlation is unity
and is plotted for the redshift range $z=2-4$, 
where we have the most confidence in the accuracy of the simulation. 
All lengths are shown in comoving units.

In order to estimate the effect of missing power at
box-sized scales, 
we computed the correlation function of the clouds
in eight subboxes. These subboxes were created by dividing the 
original simulation box into eight equal cubes and 
pretending that each subbox also has 
periodic boundary conditions.
The dotted line in Figure 15 shows the median value of the
correlation length for the subboxes, and the vertical bars
indicate the full width of the distribution at each redshift
(i.e., highest and lowest values of $r_o$).	
We see that the median correlation length drops about 30-40\%
for a box of size $L=5h^{-1}$Mpc.
Our best estimate is that the true correlation length at each 
redshift would be larger than values indicated by the solid curves
by perhaps 50\%, were the simulation box sufficiently large.

Simple extrapolation of the solid curve in Figure 15
to lower redshifts is unlikely to give
us accurate values of $r_o$ at $z\sim0.5-1.0$.
However, it is reasonable to expect the value of $r_o$ at 
these redshifts
to be significantly larger than $1.5h^{-1}$Mpc comoving,
when one takes into account the effect of the missing longer waves.
This correlation length is 
interestingly close to the observed correlation length of
the faint blue galaxies
(\cite{ebtkg91}; \cite{nwd91}; \cite{cjb93}; 
\cite{rsmf93}; \cite{ip95}; \cite{bsm95}), 
although an accurate conversion from observed angular correlation
function $\omega (\theta)$ to the 3-d correlation function as 
computed here requires detailed knowledge of the redshift distribution
of the observed faint blue objects, which is currently unavailable.
Furthermore, the average mass of these objects is close to 
$10^9-10^{10}\msun$ (see Table 1), which is in 
accord with what the mass of the faint blue 
galaxies is thought to be or required 
[\cf, e.g., Babul \& Rees' (1992) 
starburst dwarf galaxy model]. 
Finally, the number density of the clouds
(with a mean separation of $\sim 2h^{-1}$Mpc at z=2 
for $\rho_{cut}=30$) and its increasing trend with redshift
indicate that it is probably quite close to what
is thought to be for observed faint blue objects.
These three considerations lend tentative
support to the conjecture that the high redshift
$\lya$ clouds ($\nhi\ge 10^{15}$cm$^{-2}$) are the progenitors
of the faint blue galaxies.

\section{Conclusions}

We have presented a quantitative study of the sizes, shapes, 
and correlations
of $\lya$ clouds in a 
spatially flat cold dark matter universe with a cosmological 
constant, utilizing state-of-art cosmological hydrodynamic simulations
including detailed atomic physics for a 
plasma of primordial composition.
Keeping in mind the unavoidable
limitations of such numerical experiments due
to limited boxsize and limited resolution as well
as some approximate treatment of physical processes,
a few findings are probably fairly reliable.

Structures formed at high redshift $z\sim 2-4$
due to gravitational growth/collapse of cosmic density
perturbations at small-to-intermediate scales 
(100kpc to a few Mpc comoving)
are responsible for the observed $\lya$ forest.
These $\lya$ clouds
cannot be described by a simple model
which characterizes their sizes and shapes.
Their sizes vary in a wide range from a few kiloparsecs to
about one hundred kiloparsecs in proper units 
with the median size being $\sim 15-35h^{-1}$kpc at $z=3$,
and their shapes vary from nearly-spherical ellipsoids
to filaments and pancakes.
A typical $\lya$ cloud resembles a ``flattened cigar,"
with an axial ratio of $\sim 1:2:4$
at column density of $10^{13}-10^{14}$ cm$^{-2}$,
and is either a thin cigar with an axial ratios of
$(1:3:3.4)$ or near sphere with an axial ratios of $(1:1.3:1.5)$
at column density of $>10^{15}$cm$^{-2}$.
Larger clouds are, on average, less spherical than smaller ones.
The physical size of a typical cloud grows with time
roughly as $(1+z)^{-3/2}$ while its shape
hardly evolves (except for the most dense regions $\rho_{cut}>30$
where tend to be more spherical with time).

Analysis of simulated quasar double sightlines indicates
that coincident absorption features observed in the two spectra 
can have two different causes.
At small perpendicular sightline separations 
($\Delta r < 40h^{-1}$kpc proper),
a pair of coincident lines most likely represents absorption from
a common cloud.
This is usually assumed when analyzing such observations to infer
the actual cloud size.
However, at larger $\Delta r$, a pair of 
coincident lines most likely samples two separate clouds which
belong to the same cloud cluster.
In other words, clustering of separate clouds dominates the coincident 
events at larger double sightline separations.
At $\Delta r=100h^{-1}$kpc, 80-90\%
of coincident line pairs can be explained in this fashion.
At very large separations,
$\Delta r>500h^{-1}$kpc,
the coincident line events are entirely due to 
random intersections of two unrelated, uncorrelated clouds,
whose rms velocity difference $R_{||,rms}$ should relate to 
the velocity interval being examined, $D$, as $R_{||,rms} = 0.58 D$.
Analyzing observed coincident absorption lines
to infer the actual sizes of $\lya$ clouds by assuming a
population of clouds with simplified geometry (e.g. spheres, disks)
is reasonably accurate only when
the double sightline separations are small
($\Delta r \le 50h^{-1}$kpc).
For larger $\Delta r$, this exercise is not very meaningful.
The inferred sizes are not related to true sizes of the clouds
and are grossly inflated.

We attempt to make 
a connection between the high column density
$\lya$ clouds ($\nhi\ge 10^{15}$)
and the faint blue galaxies.
The supportive evidence for such a conjecture is threefold.
First, when extrapolated to $z\sim 0.5-1.0$, the correlation
 	of the high column density
$\lya$ clouds ($\ge 1.5h^{-1}$Mpc comoving)
is close to that of the observed faint blue galaxies
(\cite{ebtkg91}; \cite{nwd91}; \cite{cjb93}; 
\cite{rsmf93}; \cite{ip95}; \cite{bsm95}).
Second, the typical mass of these objects ($10^9-10^{10}\msun$)
is close to that of a dwarf galaxy [\cf, e.g.,
Babul \& Rees (1992) scenario].
Finally, the number density of the clouds
is reasonably close to what
is observed for faint blue objects.

Finally, we caution that
it may be necessary to extend the simulation's dynamic range 
to $1000-3000$ cells in each dimension,
in order to have both a larger box ($L\sim 20-30h^{-1}$Mpc comoving) and
a finer resolution ($\Delta l \sim 10-20h^{-1}$kpc comoving)
to ensure both a ``fair" sample 
and fully resolved structures of the objects in question,
while maintaining a sufficient mass resolution.

\acknowledgments
The work is supported in part
by grants NAG5-2759, AST91-08103 and ASC93-18185.
We thank the referee, George Efstathiou, 
for a constructive and pertinent report which 
improved the quality of the work.
Discussions with J.R. Gott,
N. Katz, G. Lake, J. Miralda-Escud\'e and J.P. Ostriker
are gratefully acknowledged.
RC would like to thank G. Lake and 
Department of Astronomy of University of Washington for 
the warm hospitality, and financial support 
from the NASA HPCC/ESS Program during a visit when 
much of this work was completed.

\newpage
\figcaption[FLENAME]{displays three dimensional isodensity surfaces
at $\rho_{cut}=(3,10,30)$ (in panels a,b,c), respectively, at $z=3$.
\label{fig1}}

\figcaption[FLENAME]{Figure 2a
shows two randomly selected
slices of size $5\times 5h^{-2}$Mpc$^2$
with thickness of $175h^{-1}$kpc 
at $z=3$ (all lengths are in comoving units).
The two left panels show the density contours with
$\rho_{cut} 10^{(i-1)/2}, i=1,2,3...$, and 
the two right panels show the identified clouds
in the same slices,
where two symbols are used to show two types 
of clouds: the filled dots represent ``large" clouds with their
sizes roughly indicating the actual sizes of the clouds, and
the open circles are for the ``small" clouds (see text for definitions).
Figure 2b is the same as Figure 2a
but with $\rho_{cut}=10$ for two different slices.
\label{fig2}}

\figcaption[FLENAME]{
shows the cumulative distributions of
the cloud sizes at $z=3$ for three different cases: $\rho_{cut}=(3,10,30)$,
weighted by $A$ (thin curves) and by mass (thick curves).
\label{fig3}}

\figcaption[FLENAME]{
shows the redshift evolution of the median
size of the clouds weighted by $A$ (thin curves) 
and by mass (thick curves).
Note that here the size (vertical axis) is 
in comoving length units.
\label{fig4}}

\figcaption[FLENAME]{
shows the 
distributions of the shapes of the clouds
in the $c/b$-$b/a$ plane for clouds at $z=3$
at $\rho_{cut}=3$ (5a), $\rho_{cut}=10$ (5b) and $\rho_{cut}=30$ (5c). 
Each dot represents one cloud and the contours 
indicate the density of clouds (weighted by $A$) in the plane.
The contour levels are incremented up linearly from outside to inside.
We see that the most common cloud at $\rho_{cut}=3$ and $10$
is an ellipsoid with axial ratio of $\sim 1:2:4$.
At $\rho=30$ there are two major concentrations of clouds 
at $(b/a,c/b)=(0.35,0.85)$ and $(0.75,0.85)$,
indicating the existence of two distinct 
populations: filaments and near-spherical ellipsoids.
\label{fig5}}

\figcaption[FLENAME]{
shows the shape $\eta$ as a function of size $S$ at $z=3$
for $\rho_{cut}=(3,10,30)$ (from top panel to bottom panel), respectively.
There is a trend, with large scatters, that 
larger clouds tend to be less spherical and more filamentary.
\label{fig6}}

\figcaption[FLENAME]{
represents the $A$-weighted
cumulative distribution of $\eta$ for three cases at $z=3$.
\label{fig7}}

\figcaption[FLENAME]{
shows the redshift evolution of the median shape paramter $\eta_{med}$.
\label{fig8}}

\figcaption[FLENAME]{
shows the fraction of lines which are coincident at $z=3$,
$N_{co,tot}/N_{tot}$ 
(where $N_{co,tot} = N_{co,common} + N_{co, clustering}$),
as a function of $\Delta r$.
Six curves are shown for three values of 
$\rho=(3,10,30)$ each at two values of $D=(50,150)$~km/s.
\label{fig9}}

\figcaption[FLENAME]{
shows the redshift evolution 
of $N_{co,tot}/N_{tot}$ at $\Delta r=100h^{-1}$~kpc in proper units,
for four cases with $(\nhi,D)$ of
$(3\times 10^{13},150)$,
$(3\times 10^{13},50)$,
$(3\times 10^{14},150)$,
$(3\times 10^{14},50)$.
\label{fig10}}

\figcaption[FLENAME]{
shows the ratio of coincident line pairs,
each of which shares the same cloud,
to the total number of coincident line pairs, $N_{co,common}/N_{co,tot}$,
at $z=3$.  Six curves are shown 
for $\rho=(3,10,30)$ with $D=50$~km/s and $150$~km/s.
\label{fig11}}

\figcaption[FLENAME]{
shows rms difference of coincident line pairs along the
line of sight, $R_{||,rms}$, as a function of the sigthline separation
perpendicular to the line of sight, $\Delta r$,
 for $D=50km/s$ and 150km/s at $\rho_{cut}=(3,10,30)$. 
Also shown as the two long-dashed horizontal lines are the rms separations
for the two $D$ values, if clouds are small (compared 
to perpendicular separation) and randomly distributed.
\label{fig12}}

\figcaption[FLENAME]{
shows the redshift evolution of $N_{co, common}/N_{co,tot}$
for $\nhi=3\times 10^{13}$cm$^{-2}$ (panel a)
and $\nhi=3\times 10^{14}$cm$^{-2}$ (panel b), respectively.
Four cases are shown in each figure
with two separations, $\Delta r=50h^{-1}$kpc and 
$\Delta r=100h^{-1}$kpc both in proper units, and
with $D=50$km/s and $150$km/s.
\label{fig13}}

\figcaption[FLENAME]{
shows the correlation of  $\lya$ clouds at $z=3$
for $\rho_{cut}=(10,30)$.
\label{fig14}}

\figcaption[FLENAME]{
shows the redshift evolution of 
correlation length $r_o$ (solid curve)
defined as the length where correlation is unity
(in comoving length units).
Also shown as the dotted curve is the correlation length
computed using the eight subboxes to illustrate 
the effect of the simulation boxsize on the correlation.
\label{fig15}}

\clearpage
\begin{deluxetable}{cccccccc} %{l,r}
%\small
%\footnotesize
%\scriptsize
\tablewidth{0pt}
\tablenum{1}
\tablecolumns{8}
\tablecaption{Cloud Statistics} %\label{tab1}}
\tablehead{
\colhead{Redshift} &
\colhead{$\rho_{cut}$} &
\colhead{\# of Clouds} &
\colhead{Mass fraction} &
\colhead{$\langle r_{sp}\rangle$ (Mpc proper)}&
\colhead{$\langle m_{cloud}\rangle$ ($\msun$)}}

%\startdata
%$2$ & $3.0$ & $1759$ & $0.51$ & $0.28$ & $2.8\times 10^{9}$ \nl
%$3$ & $3.0$ & $2639$ & $0.43$ & $0.18$ & $1.6\times 10^{9}$ \nl
%$4$ & $3.0$ & $3421$ & $0.36$ & $0.13$ & $1.0\times 10^{9}$ \nl
%\hline
%$2$ & $10.0$ & $1064$ & $0.33$ & $0.33$ & $3.0\times 10^{9}$ \nl
%$3$ & $10.0$ & $1548$ & $0.25$ & $0.22$ & $1.5\times 10^{9}$ \nl
%$4$ & $10.0$ & $1952$ & $0.19$ & $0.16$ & $9.0\times 10^{8}$ \nl
%\hline
%$2$ & $30.0$  & $639$ & $0.20$ & $0.39$ & $3.1\times 10^{9}$ \nl
%$3$ & $30.0$ & $926$ & $0.14$  & $0.26$  & $1.4\times 10^{9}$ \nl
%$4$ & $30.0$  & $1117$ & $0.10$ & $0.19$ & $8.2\times 10^{8}$ \nl
%\enddata

\startdata
$2$ & $3.0$ & $2027$ & $0.51$ & $0.42$ & $3.8\times 10^{9}$ \nl
$3$ & $3.0$ & $3025$ & $0.43$ & $0.26$ & $2.1\times 10^{9}$ \nl
$4$ & $3.0$ & $3955$ & $0.36$ & $0.20$ & $1.3\times 10^{9}$ \nl
\hline
$2$ & $10.0$ & $1191$ & $0.33$ & $0.49$ & $4.1\times 10^{9}$ \nl
$3$ & $10.0$ & $1696$ & $0.25$ & $0.32$ & $2.1\times 10^{9}$ \nl
$4$ & $10.0$ & $2197$ & $0.19$ & $0.23$ & $1.2\times 10^{9}$ \nl
\hline
$2$ & $30.0$ & $683$  & $0.20$ & $0.57$ & $4.3\times 10^{9}$ \nl
$3$ & $30.0$ & $986$  & $0.14$ & $0.38$ & $2.0\times 10^{9}$ \nl
$4$ & $30.0$ & $1242$ & $0.10$ & $0.28$ & $1.1\times 10^{9}$ \nl
\enddata
%\tablenotetext{a}{add}
%\tablecomment{TEXT}
%\tablerefs{TEXT}
\end{deluxetable}


\begin{thebibliography}{DUM}
\bibitem[Adams \etal 1993]{a93} Adams, F., Bond, J.R., Freese, K., Frieman, J. \& Olinto, A., 1993, preprint
\bibitem[Babul \& Rees 1992]{br92} Babul, A., \& Rees, M. 1992, \mnras, 255, 346
\bibitem[Bahcall \& Cen 1992]{bc92} Bahcall, N., \& Cen, R. 1992, \aplett, 407, L49 
\bibitem[Bahcall \& Cen 1993]{bc93} Bahcall, N.A., \& Cen, R. 1993, \aplett, 407, L49 
\bibitem[Bechtold \etal 1994]{b94} Bechtold, J., Crotts, A.P.S., Duncan, R.C., Fang, Y. 1994, \apj, 437, L83
\bibitem[Bechtold \& Yee 1996]{by96} Bechtold, J., \& Yee, H.K.C. 1996, ApJ, in press
\bibitem[Bertschinger \& Gelb 1991]{bg91} Bertschinger, E., \& Gelb, J.M. 1991, Comp. Phys., 5, 164 
\bibitem[Bolte \& Hogan 1995]{bh95} Bolte, M., \& Hogan, C.J. 1995, \nat, 376, 399
\bibitem[Bond \& Myers 1996]{bm96} Bond, J.R., Myers, 1996, \apj, 
\bibitem[Brainerd, Smail, \& Mould 1995]{bsm95} Brainerd, T.G., Smail, I, \& Mould, J. 1995, preprint
\bibitem[Bucher, Goldhaber, \& Turok 1995]{bgt95} Bucher, Goldhaber, \& Turok, N. 1995, preprint
\bibitem[Carswell \etal 1991]{c91} Carswell, R. F., Lanzetta, K. M., Parnell, H. C., \& Webb, J. K., 1991, \apj, 371, 36
\bibitem[Cen 1992]{c92} Cen, R. 1992, \apjs, 78, 341
\bibitem[Cen \& Ostriker 1993a]{co93a} Cen, R., \& Ostriker, J.P. 1993a, \apj, 417, 404 
\bibitem[Cen \& Ostriker 1993b]{co93b} Cen, R., \& Ostriker, J.P. 1993b, \apj, 417, 415 
\bibitem[Cen \& Ostriker 1994]{co94} Cen, R., \& Ostriker, J.P. 1994, \apj, 431, 451
\bibitem[Cen, Gnedin, \& Ostriker 1993]{cgo93} Cen, R., Gnedin, N.Y., \& Ostriker, J.P. 1993, \apj, 417, 387
\bibitem[Cen \etal 1992]{cgko92} Cen, R., Gnedin, N.Y., Kofman, L.A., \& Ostriker, J.P. 1992, \apj, 399, L11
\bibitem[Cen, Ostriker, \& Peebles 1993] {cop93} Cen, R., Ostriker, J.P., \& Peebles, P.J.E. 1993, \apj, 415, 423 
\bibitem[Cen \etal 1994b]{cmor94} Cen, R., Miralda-Escud\'e, J., Ostriker, J.~P., \& Rauch, M. 1994, \apj, 437, L9
\bibitem[Couch, Jurcevic, \& Boyle 1993]{cjb93} Couch, W.J., Jurcevic, J.S., \& Boyle, B.J. 1993, \mnras, 260, 241
\bibitem[Cowie \etal 1988]{clgm88} Cowie, L.L., Lilly, S.J., Gardner, J.P., \& McLean, I.S. 1988, \apj, 332, L29
\bibitem[Cristiani \etal 1995]{c95} Cristiani, S., D'Odorico, S., Fontana, A., Giallongo, E., \& Savaglio, S. 1995, \mnras, 273, 1016
\bibitem[Davis, Summers \& Schlegel 1992]{dss92} Davis, M., Summers, F.J., \& Schlegel, D. 1992, \nat, 359, 393 
\bibitem[Dinshaw \etal 1994]{d94} Dinshaw, N., Impey, C.D., Foltz, C.B., Weymann, R.J., \& Chaffee, F.H. 1994, \apj, 437, L87
\bibitem[Dinshaw \etal 1995]{d95} Dinshaw, N., Foltz, C.D., Impey, C.D., Weymann, R.J., \& Morris, S.L. 1995, \nat, 373, 223
\bibitem[Edelson \& Malkan 1986]{em86} Edelson, \& Malkan 1986, \apj, 308, 59
\bibitem[Efstathiou, Bond, \& White 1992]{ebw92} Efstathiou, G., Bond, J.R., \& White, S.D.M. 1992, \mnras, 258, 1p
\bibitem[Efstathiou \etal 1991]{ebtkg91} Efstathiou, G., Bernstein, G., Tyson, J.A., Katz, N., \& Guhathakurta, P. 1991, \apj, 380, L47
\bibitem[Eke, Cole, \& Frenk 1996]{ecf96} Eke, V.R., Cole, S., \& Frenk, C.S. 1996, preprint
\bibitem[Fang \etal 1996]{fdcb96} Fang, Y., Dunca, R.C., Crotts, A.P.S., \& Bechtold, J. 1996, preprint
\bibitem[Feldman, Kaiser, \& Peacock 1994]{fnp94} Feldman, H.A., Kaiser, N., \& Peacock, J.A. 1994, \apj, 437, 56
\bibitem[Freedman \etal 1994]{f94} Freedman, M.L., \etal 1994, \nat, 371, 757
\bibitem[Gott 1982]{g82} Gott, J.R., III 1982, \nat, 295, 304
\bibitem[Hamuy \etal 1995]{m95} Hamuy, M., \etal 1995, \aj, 109, 1
\bibitem[Hernquist, Katz, \& Weinberg 1996]{hkw96} Hernquist, L., Katz, N., \& Weinberg, D.H. 1996, \apjl, 457, L51
\bibitem[Hu \etal 1995]{h95} Hu, E. M., Kim, T.-S., Cowie, L. L., Songaila, A., \& Rauch, M. 1995, \aj, 110, 1526
\bibitem[Infante \& Pritchet 1995]{ip95} Infante, L., \& Pritchet, C.J. 1995, \apj, 439, 565
\bibitem[Klypin \etal 1993]{k93} Klypin, A., Holtzman, J., Primack, J., \& Regos, E. 1993, \apj, 416, 1
\bibitem[Kofman, Gnedin, \& Bahcall 1993]{kgb93} Kofman, L.A, Gnedin, N.Y., \& Bahcall, N.A. 1993, \apj, 413, 1
\bibitem[Koo 1986]{k86} Koo, D. 1986, \apj, 311, 651
\bibitem[Liddle, Lyth, \& Sutherland 1992]{lls92} Liddle, A.R., Lyth, D.H., \& Sutherland, W. 1992, Phys. Lett., B 279, 244
\bibitem[Lidsey \& Coles 1992]{lc92} Lidsey, J.E., Coles, P. 1992, \mnras, 258, 57p
\bibitem[Lucchin, Matarrese, \& Mollerach 1993]{lmm93} Lucchin, F., Matarrese, S., \& Mollerach, S., 1993, submitted to ApJ(Letters)
\bibitem[Ma \& Bertschinger 1994]{mb94} Ma, C.P., \& Bertschinger, E. 1994, \apj, 435, L9
\bibitem[Miralda-Escud\'e \etal 1996]{mcor95} Miralda-Escud\'e, J., Cen, R., Ostriker, J.~P., \& Rauch, M. 1996, \apj, in press
\bibitem[Neuschaefer, Windhorst, \& Dressler 1991]{nwd91} Neuschaefer, L.W., Windhorst, R.A., \& Dressler, A. 1991, \apj, 382, 32
\bibitem[Ostriker \& Steinhardt 1995]{os95} Ostriker, J.P., \& Steinhardt, P. 1995, \nat, 377, 600
\bibitem[Oukbir \& Blanchard 1992]{ob92} Oukbir, J., \& Blanchard, A. 1992, A\& A, 262, L21
\bibitem[Peacock, \& Dodd 1994]{pd94} Peacock, J.A., \& Dodds, S.J. 1994, \mnras, 267, 1020
\bibitem[Peebles 1984]{p84} Peebles, P.J.E. 1984, ApJ, 284, 439
\bibitem[Peebles 1987]{p87a} Peeles, P.J.E 1987a, \nat, 327, 210
\bibitem[Peebles 1987]{p87b} Peeles, P.J.E 1987b, \apj, 315, L73
\bibitem[Pen 1996]{p96} Pen, U.-L. 1996, in preparation
\bibitem[Petitjean \etal 1993]{p93} Petitjean, P., Webb, J. K., Rauch. M., Carswell, R. F., \& Lanzetta, K., 1993, \mnras, 262, 499
\bibitem[Rauch \etal 1992]{r92} Rauch, M., Carswell, R. F, Chaffee, F. H., Foltz, C. B., Webb, J. K., Weymann, R. J., Bechtold, J., \& Green, R. F. 1992, \apj, 390, 387
\bibitem[Riess, Press, \& Kirshner 1995]{rpk95} Riess, A.G., Press, W.H., \& Kirshner, R.P. 1995, \apj, 438, L17
\bibitem[Roche \etal 1993]{rsmf93} Roche, N., Shanks, T., Metcalfe, N., \& Fong, R. 1993, \mnras, 263, 360
\bibitem[Scalo 1986]{s86} Scalo, J.M. 1986, Fund. Cosmic Phys., 11, 1
\bibitem[Schneider \etal 1993]{s93} Schneider, D.P., et al. 1993, \apjs, 87, 45
\bibitem[Smette \etal 1992]{sssfcwwm92} Smette, A., Surdej, J., Shaver, P.A., Foltz, C.B., Chaffee, F.H., Weymann, R.J., Williams, R.E., \& Magain, P. 1992, \apj, 389, 39
\bibitem[Smette \etal 1995]{srsrwk95} Smette, A., Robertson, J.G., Shaver, P.A., Reimers, D., Wisotzki, L., \& Kohler, T.H. 1995, in press
\bibitem[Smoot \etal 1992]{s92} Smoot, G.F., \etal 1992, \aplett, 396, L1
\bibitem[Taylor \& Rowan-Robinson 1992]{tr92} Taylor, A.N., \& Rowan-Robinson, M. 1992, \nat, 359, 396
\bibitem[Tyson 1988]{t88} Tyson, J.A. 1988, AJ, 96, 1
\bibitem[Tytler, Fan, \& Burles 1996]{tfb96} Tytler, D., Fan, X.-M., \& Burles, S. 1996, \nat, 381, 207
\bibitem[Tytler \etal 1995]{t95} Tytler, D., Fan, X.-M., Burles, S., Cottrell, L., Davis, C., Kirkman, D., \& Zuo, L. 1995, in {\it QSO Absorption Lines}, Proc. ESO Workshop, ed. G. Meylan (Heidelberg: Springer), p. 289
\bibitem[Viana \& Liddle 1995]{vl95} Viana, P.T.P, \& Liddle, A.R. 1995, preprint
\bibitem[Walker \etal 1991]{w91} Walker, T.P., Steigman, G., Schramm, D.N., Olive, K.A., \& Kang, H.S. 1991, \apj, 376, 51
%\bibitem[Wambsganss \etal 1995]{wcot95} Wambsganss, J., Cen, R., Ostriker, J.P., \& Turner, E.L. 1995, Science, 268, 274
\bibitem[White \etal 1993a]{wef93} White, S.D.M., Efstathiou, G., \& Frenk, C.S. 1993, \mnras, 262, 1023
\bibitem[White \etal 1993b]{wnef93} White, S.D.M., Navarro, J.F., Evrard, A.E., \& Frenk, C.S. 1993, \nat, 366, 429
\bibitem[Zhang \etal 1995]{zan95} Zhang, Y., Anninos, P., \& Norman, M.L. 1995, \apjl, 453, L57
\end{thebibliography}
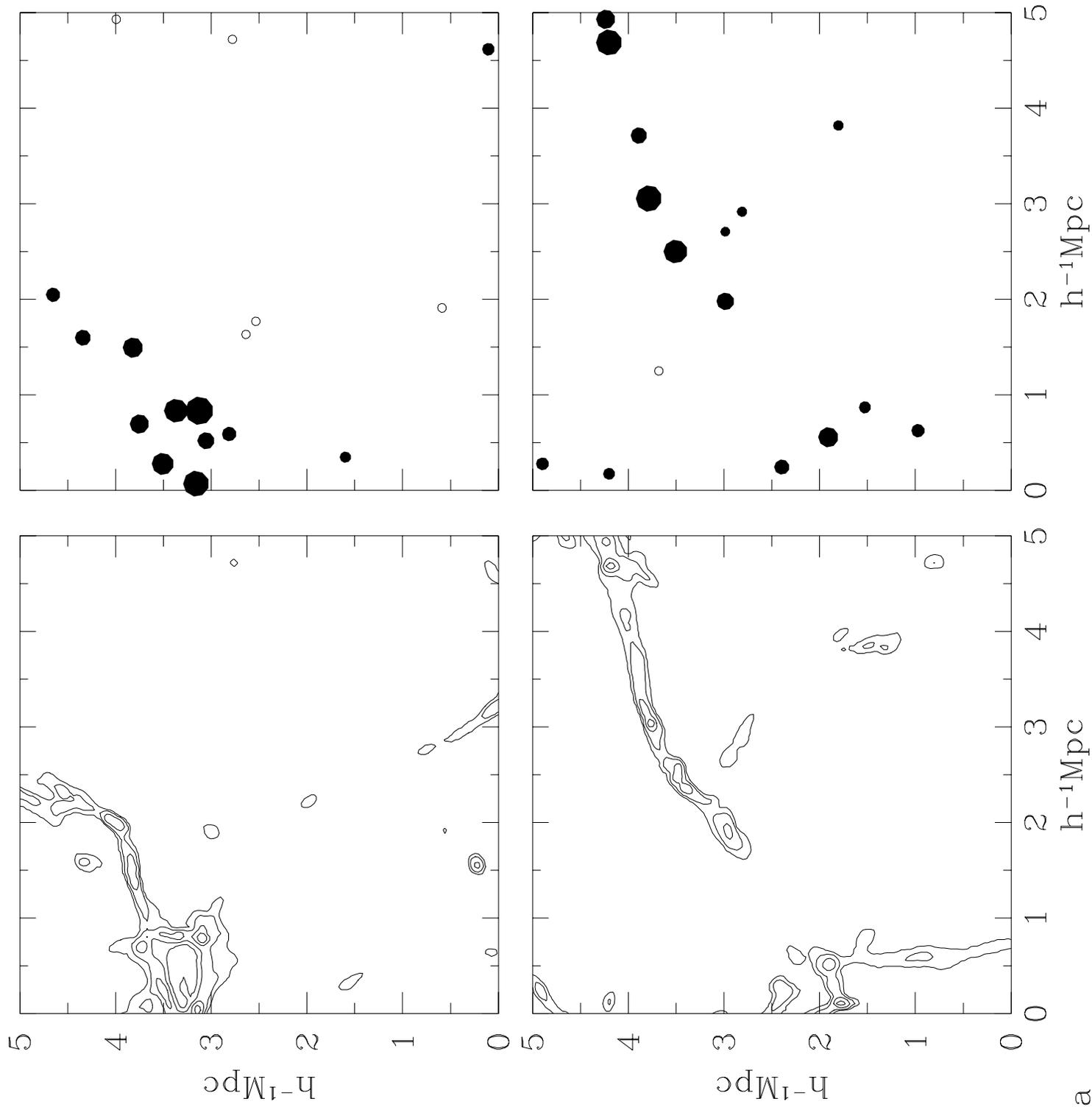
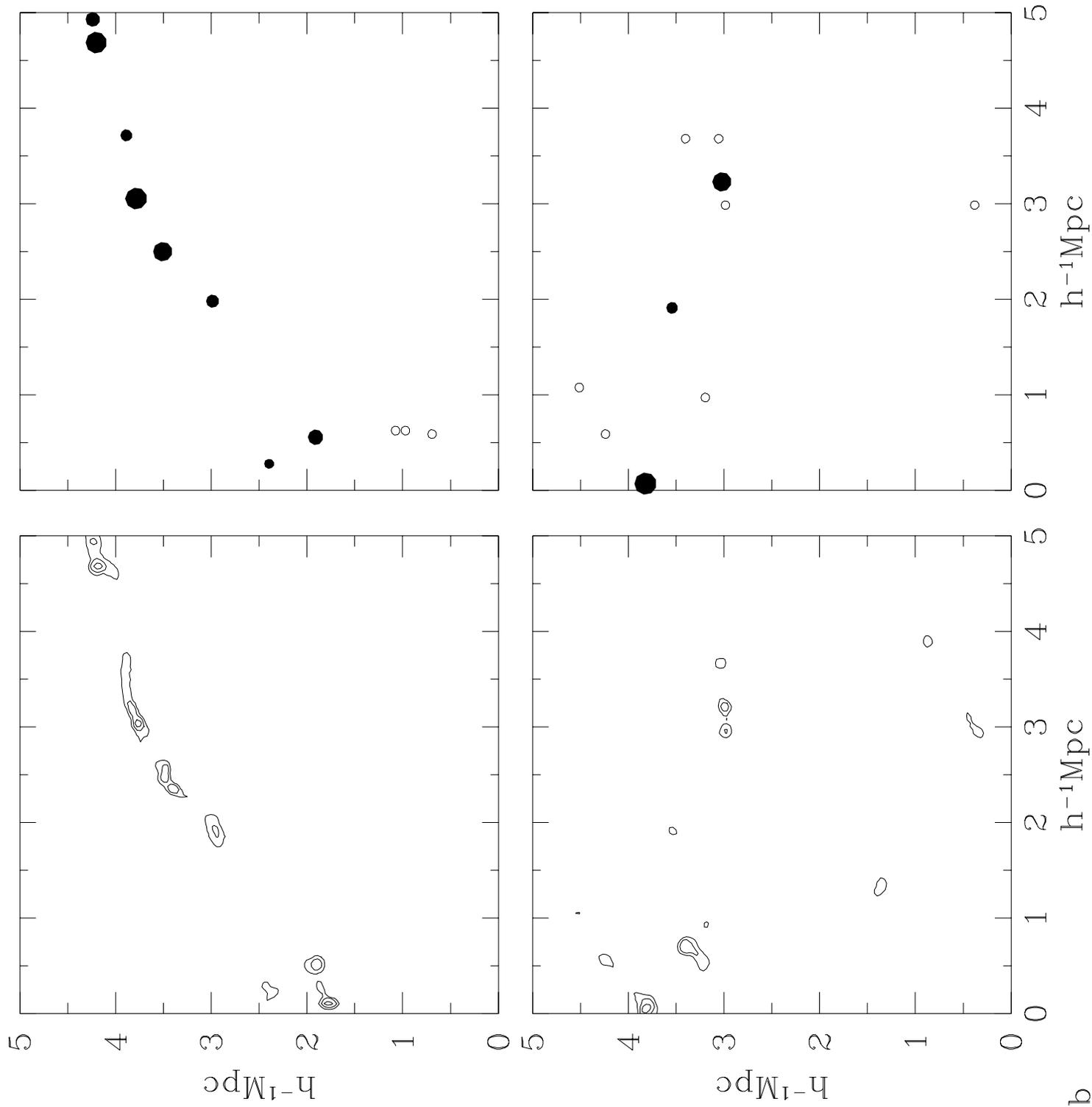
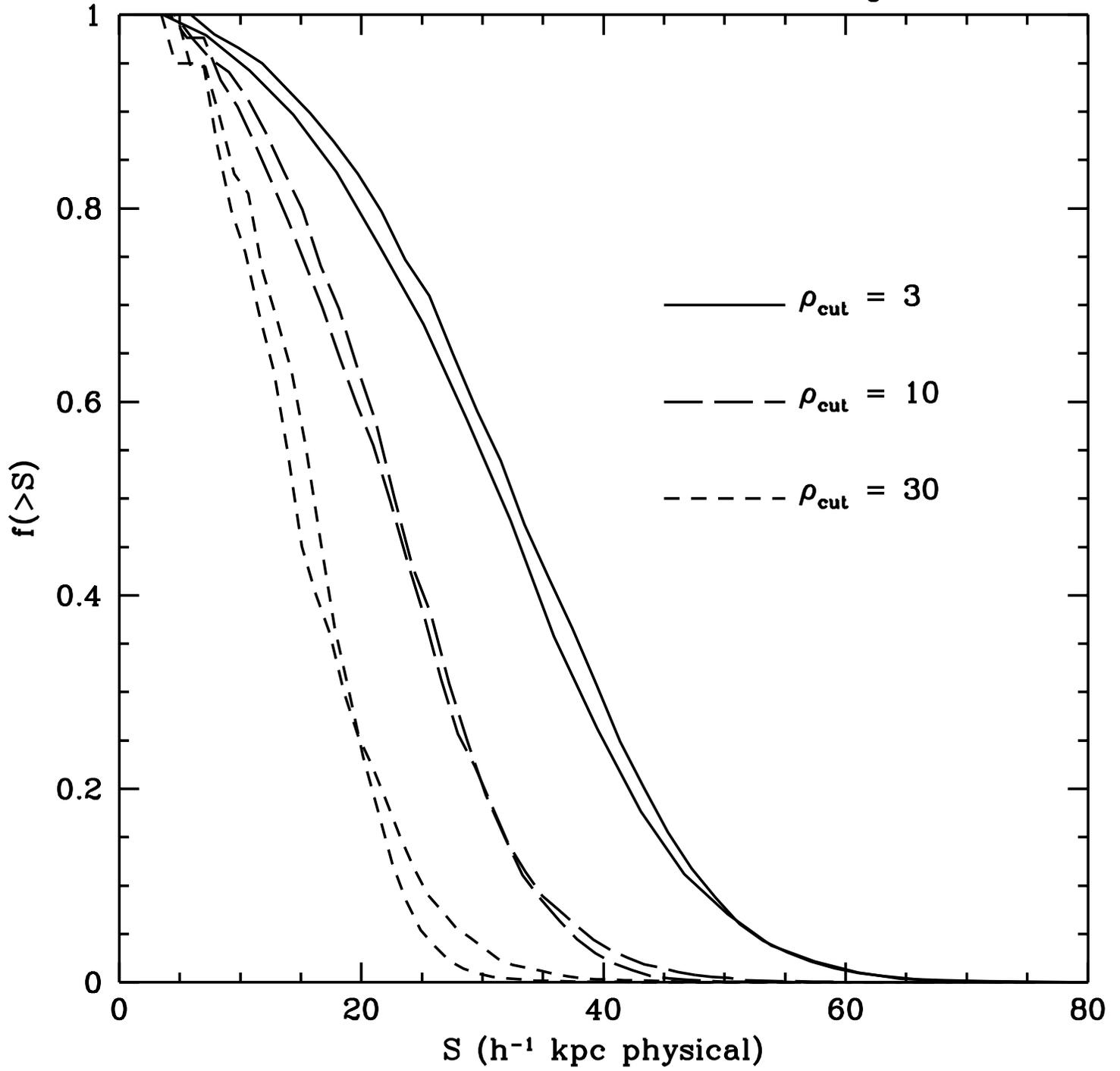
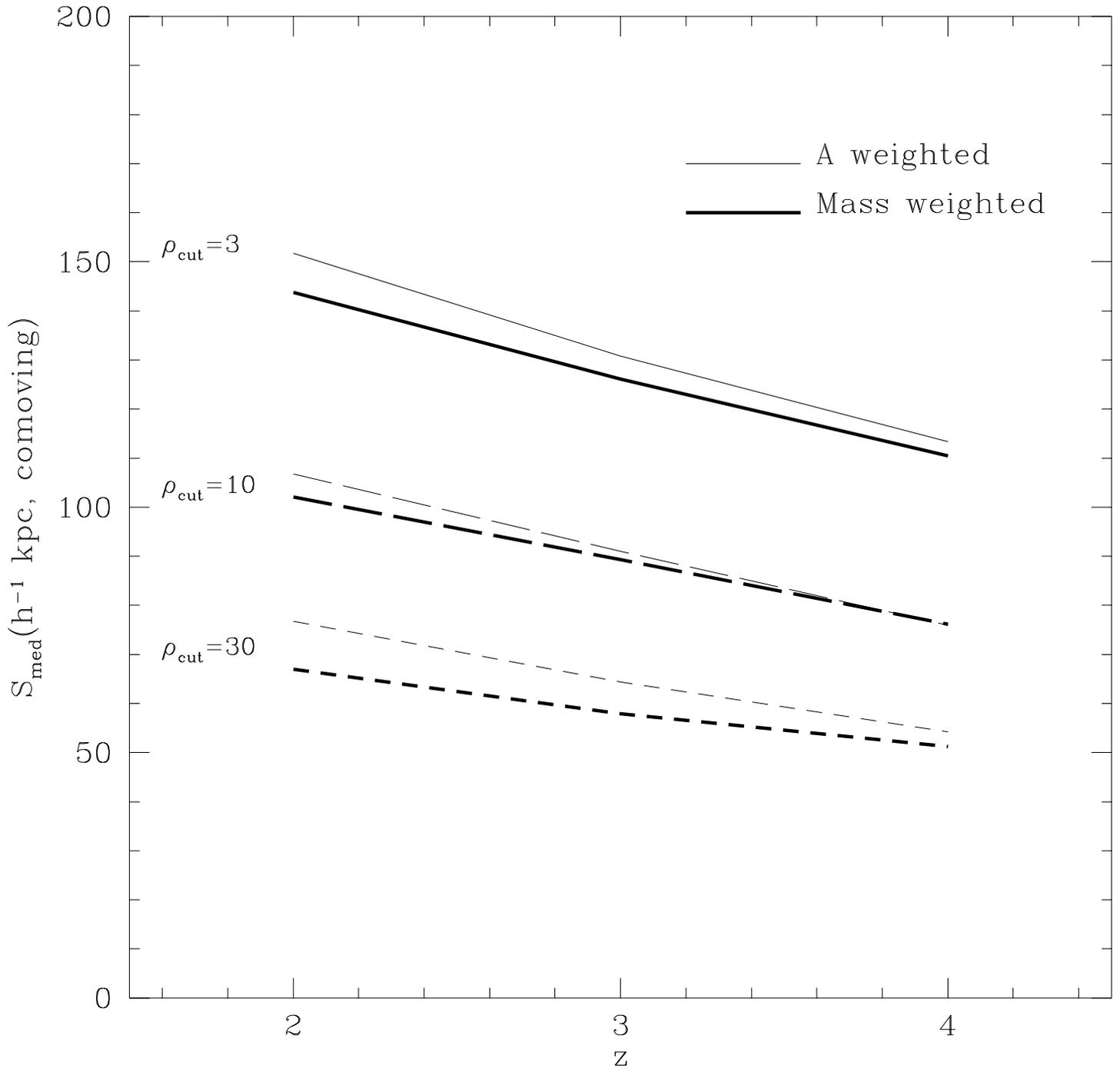
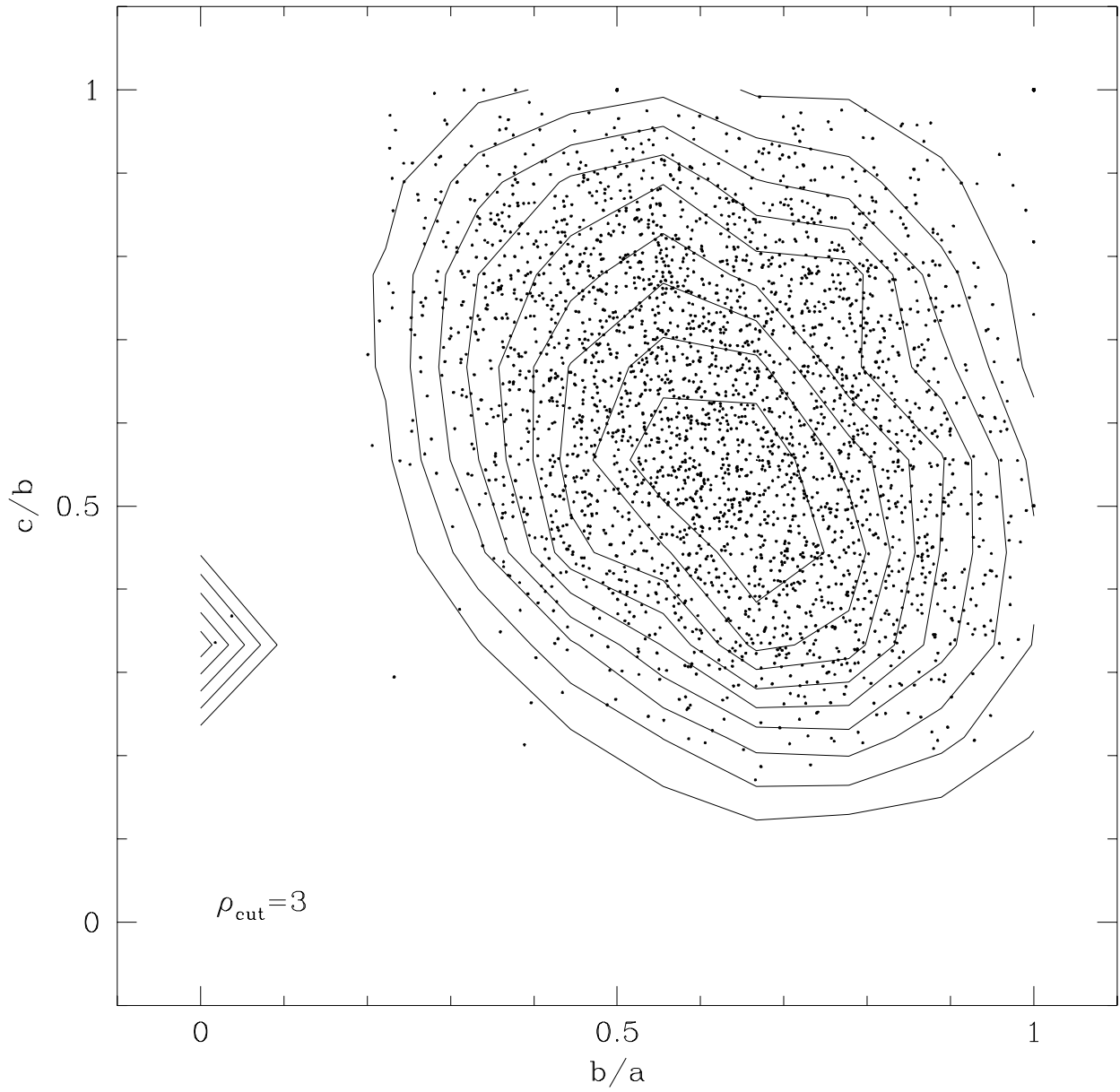
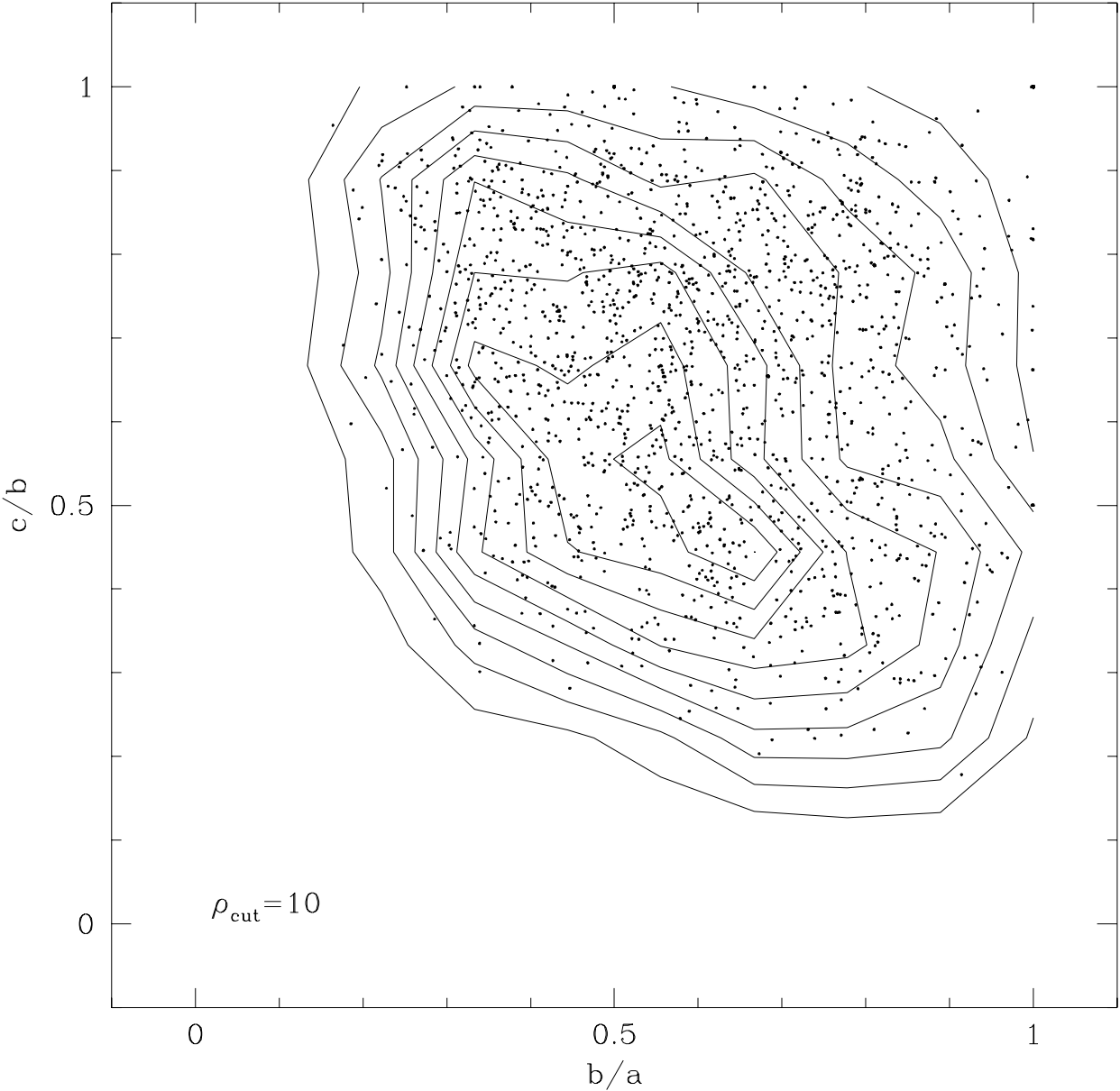
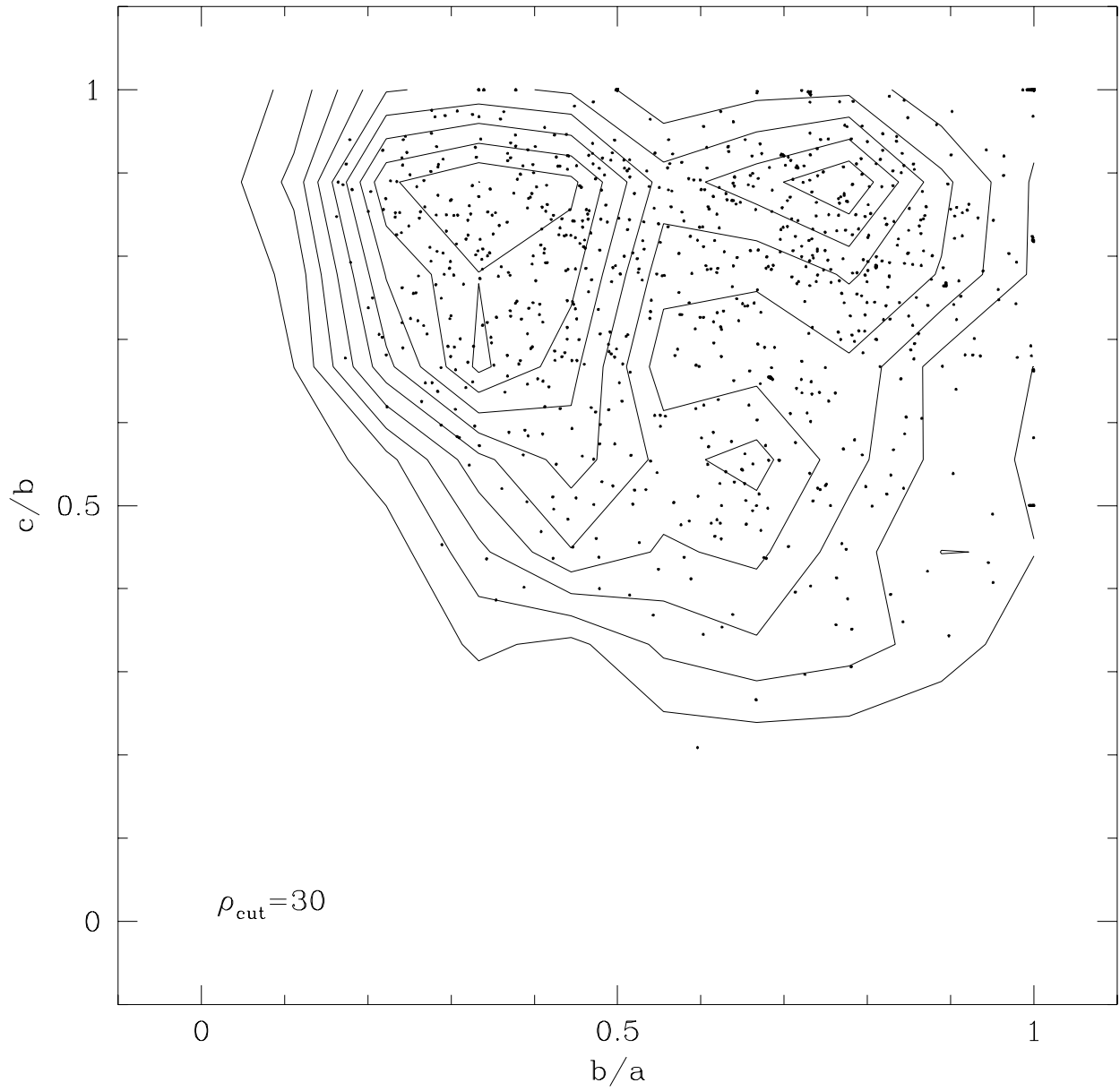
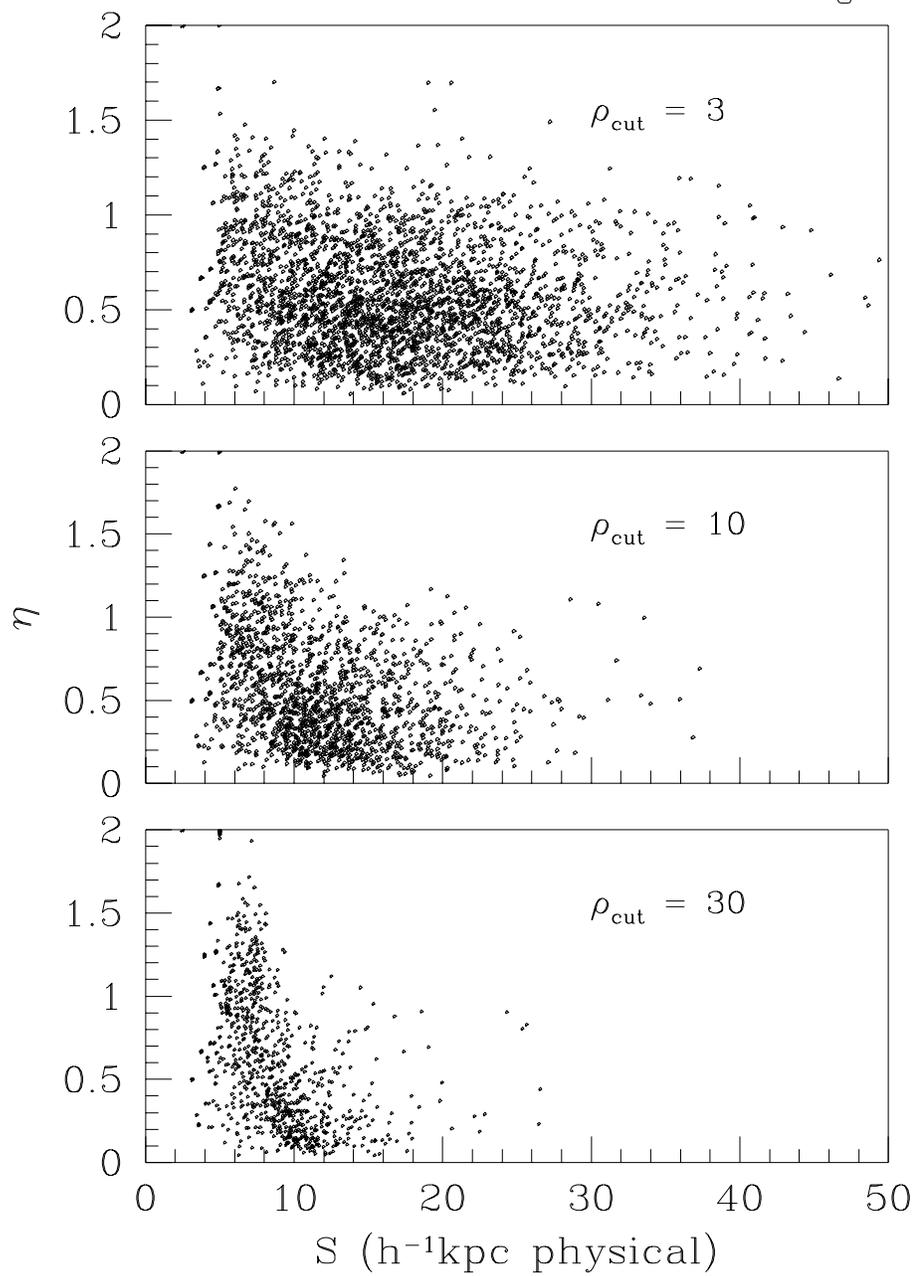
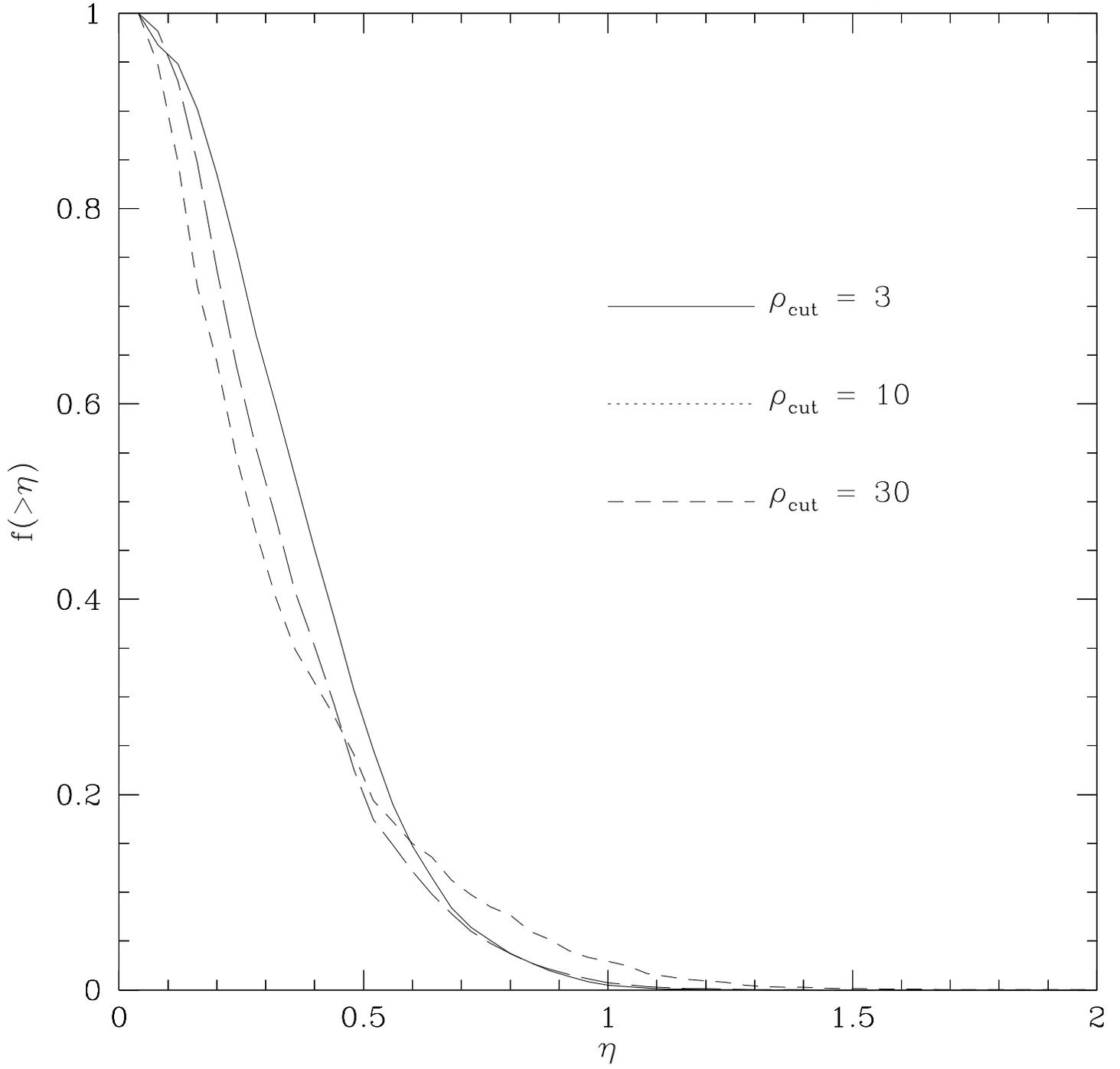
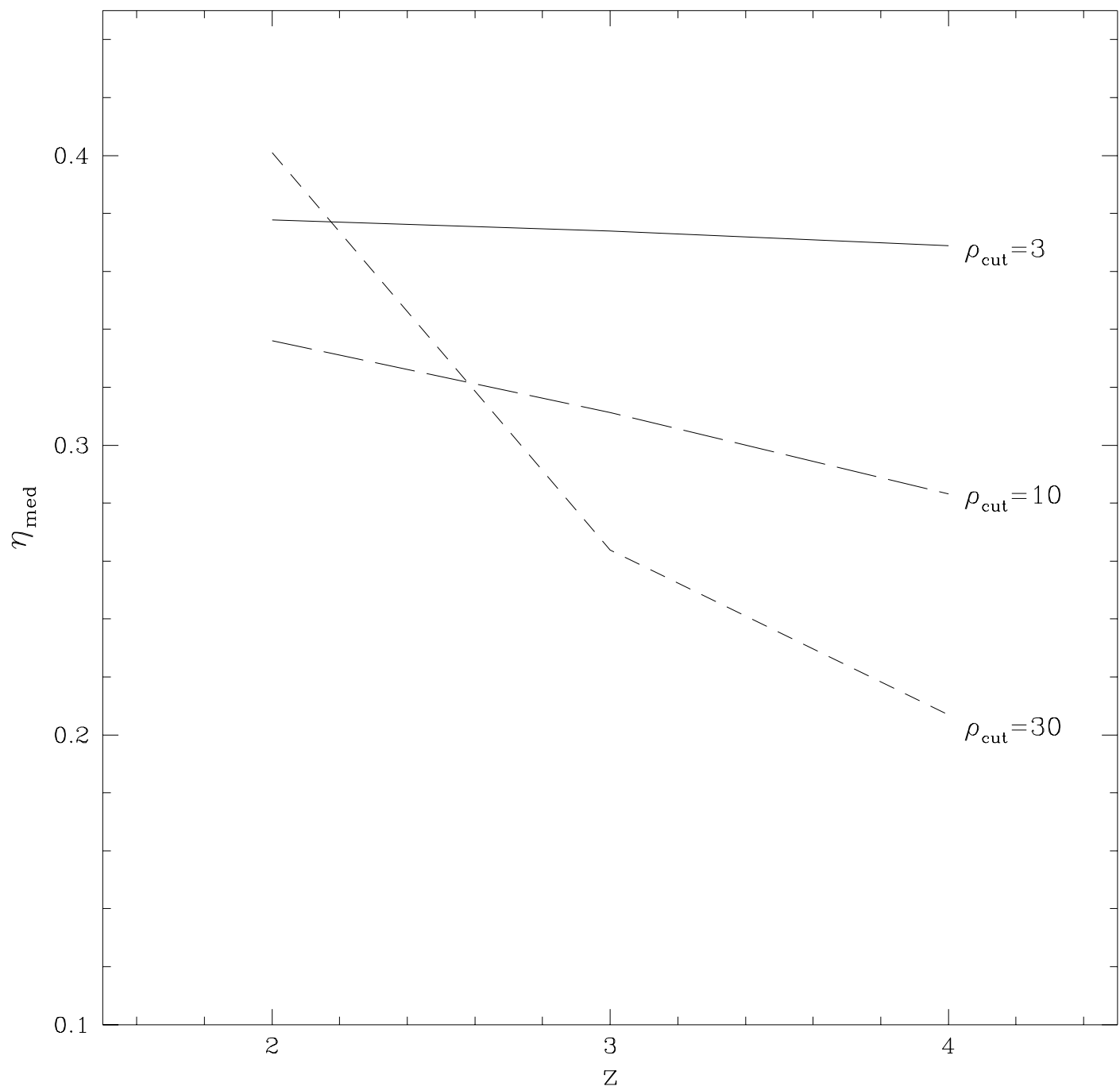
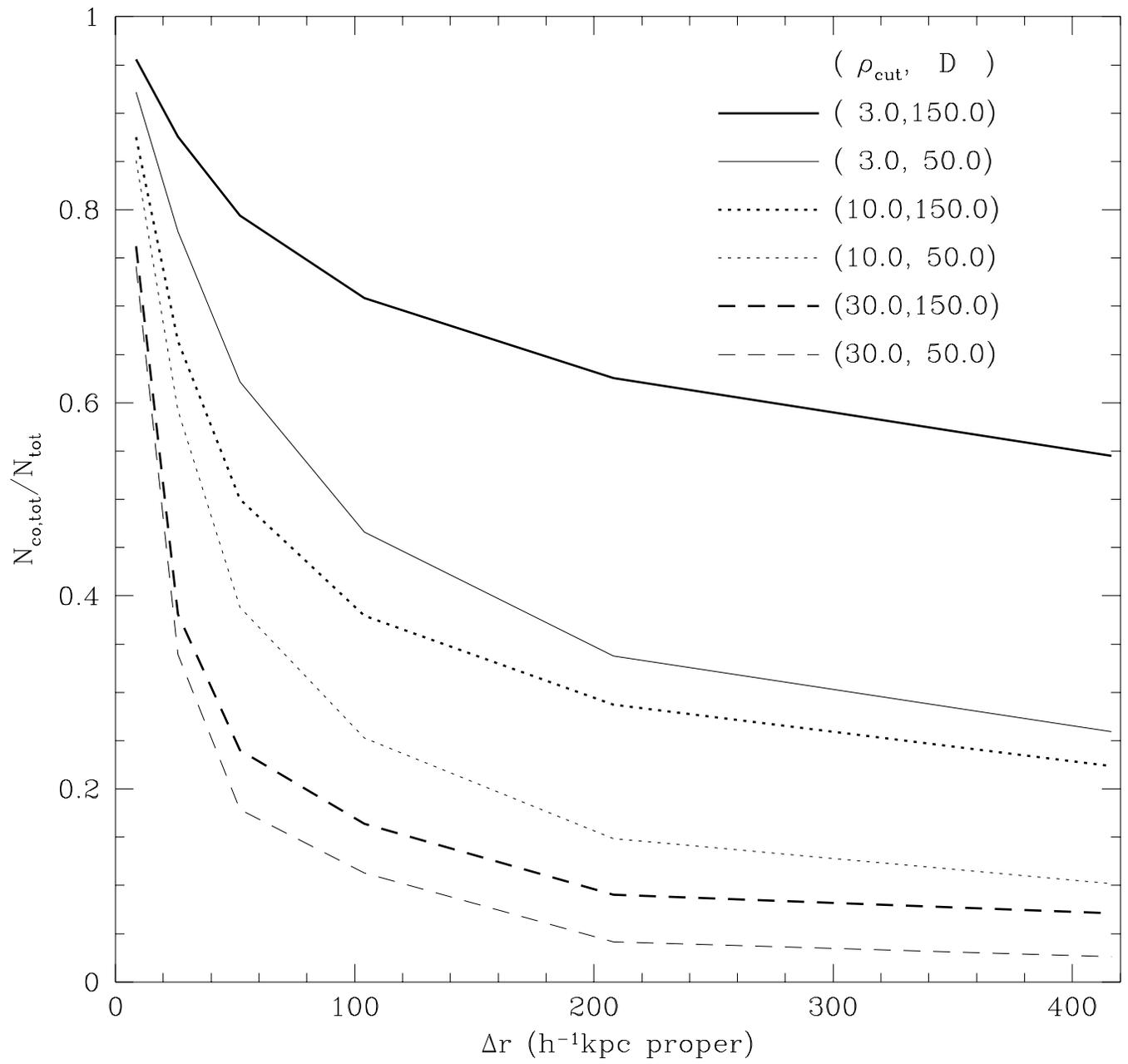
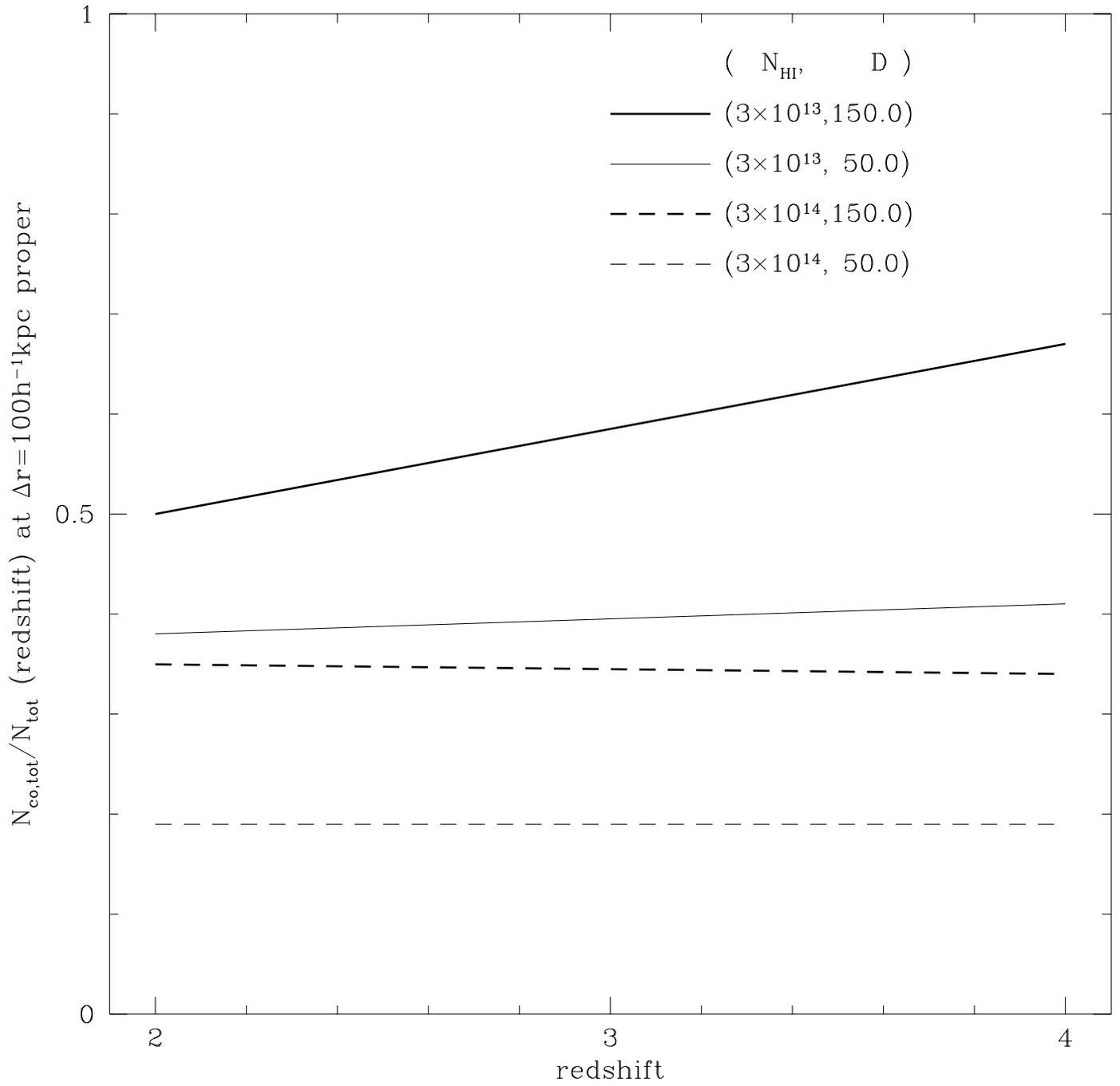
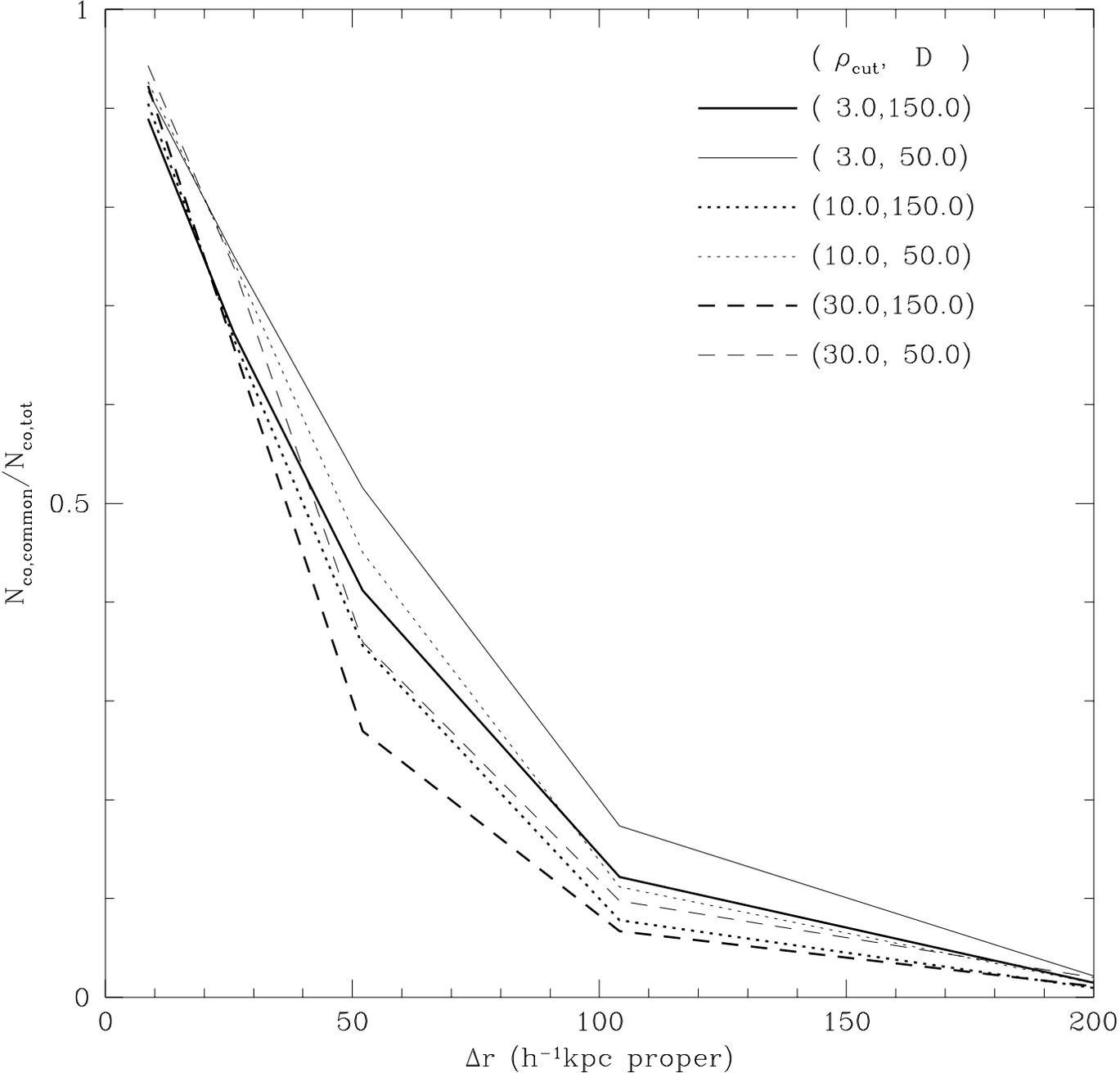
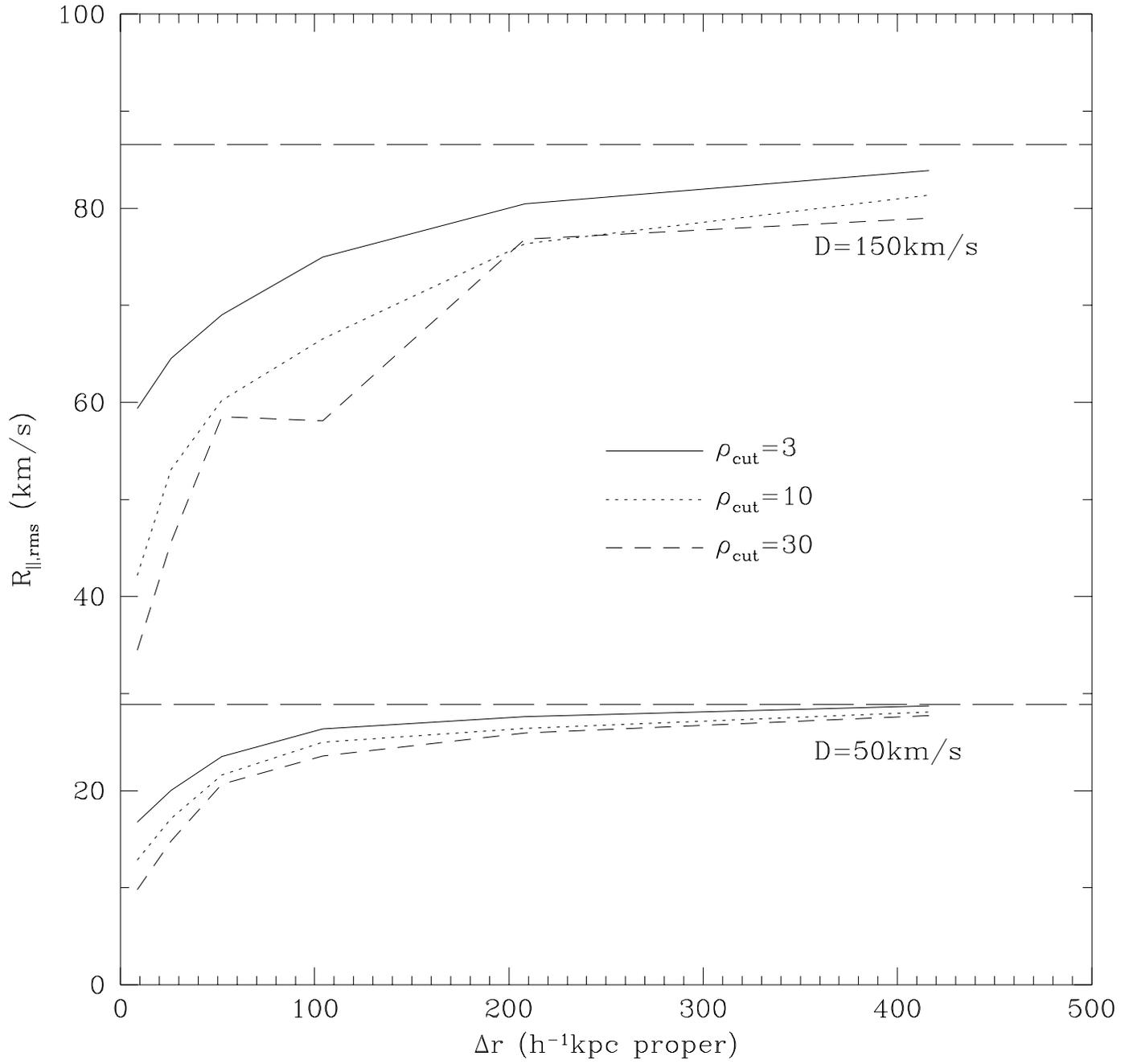
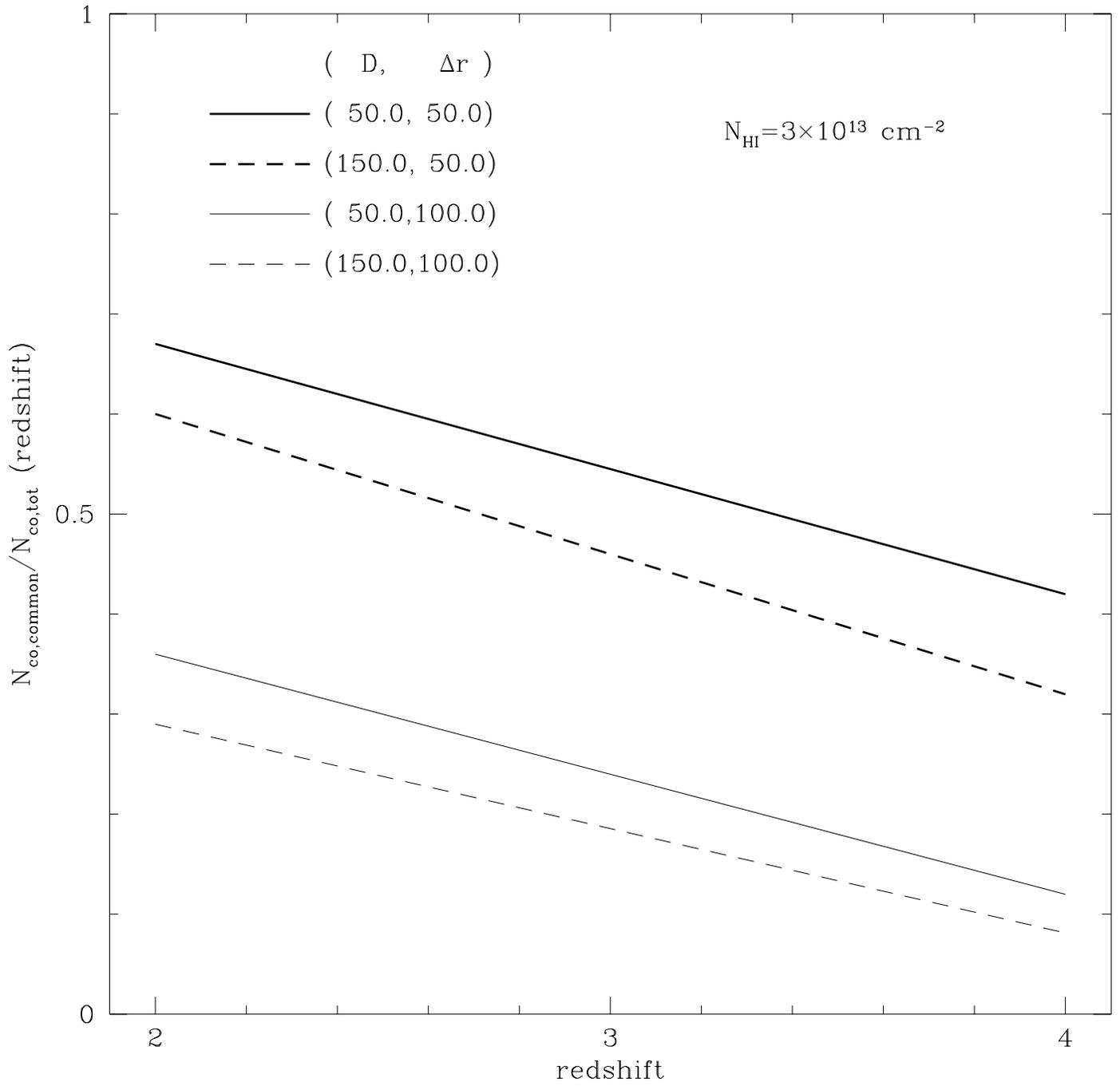
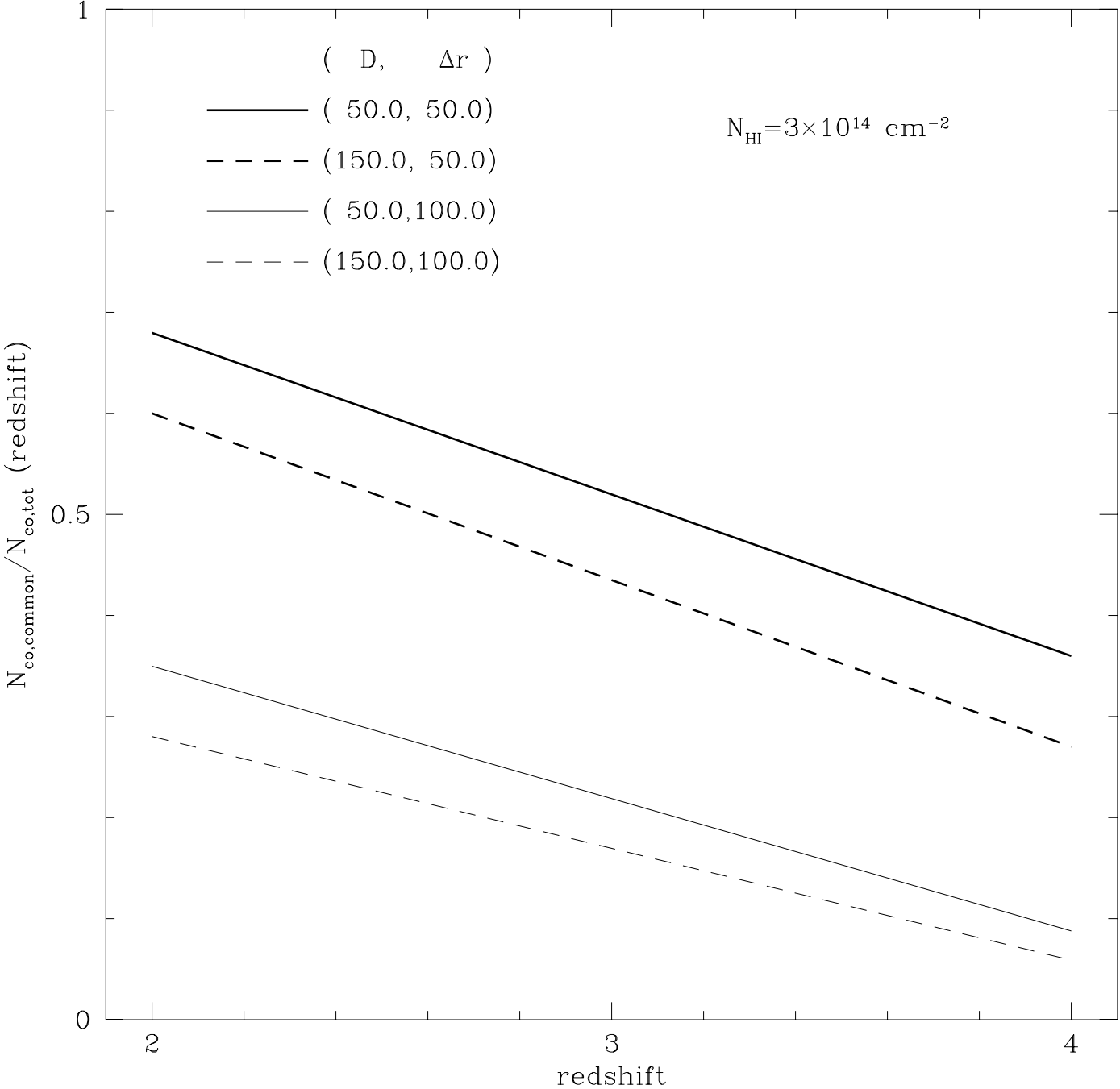
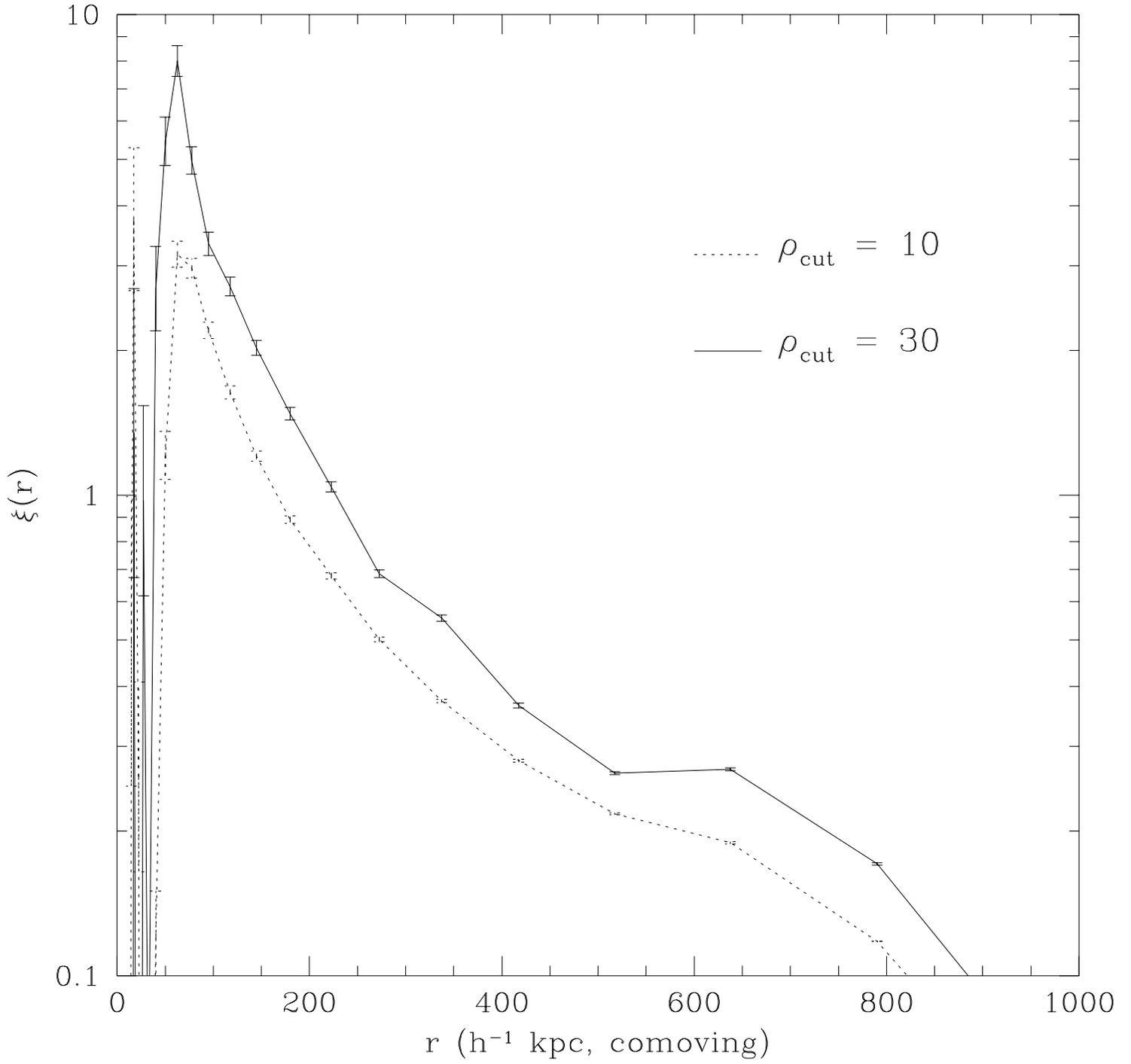
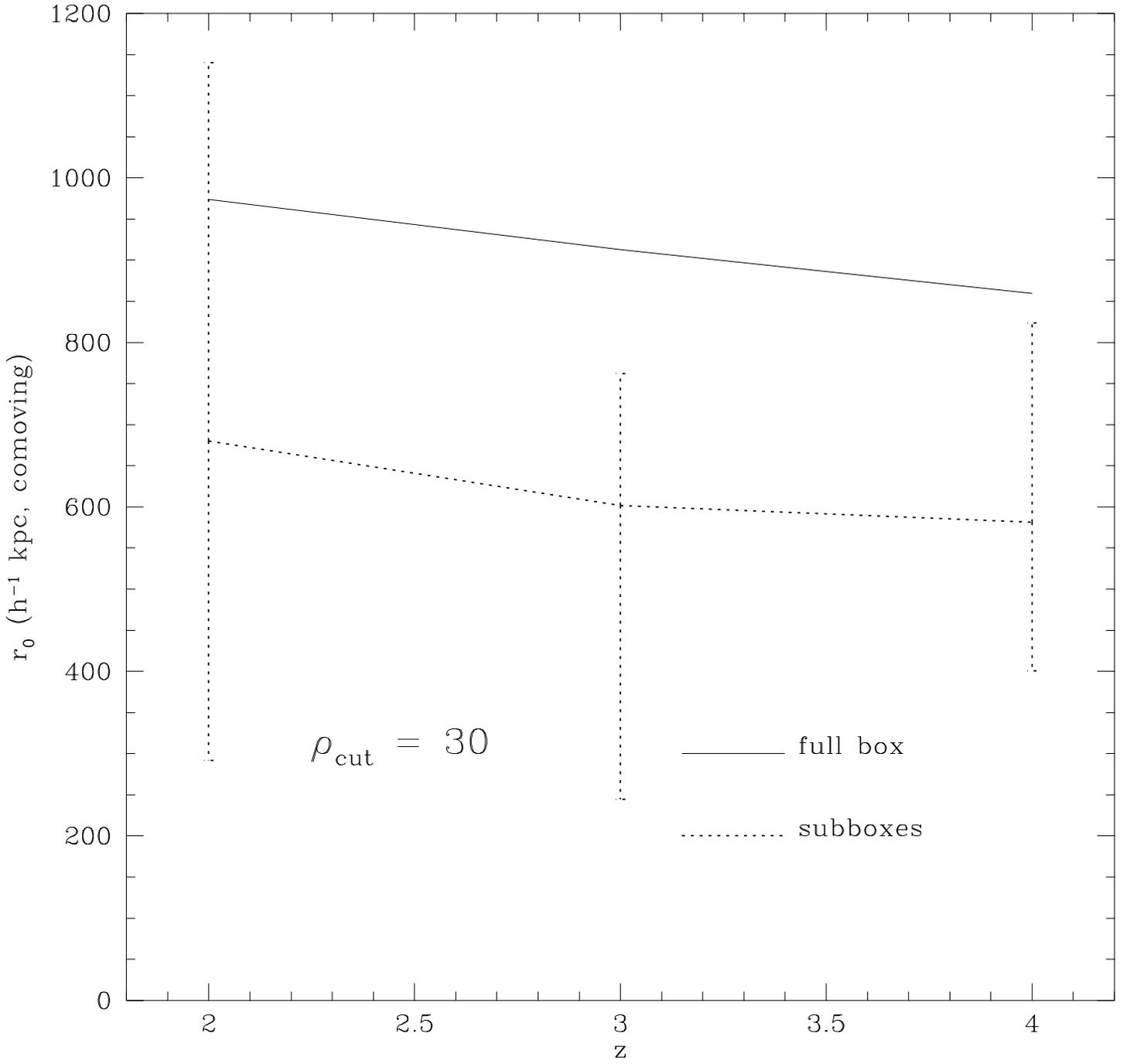
\end{document}